\documentclass[aps,prd,preprint,groupedaddress,showkeys,showpacs,amsmath]
              {revtex4}

\newcommand{\ele}{{\mathrm e}}
\newcommand{\neu}{{\mathrm n}}
\newcommand{\pro}{{\mathrm p}}
\newcommand{\deu}{{\mathrm d}}

\newcommand{\he}{{\mathrm H}{\mathrm e}}
\newcommand{\li}{{\mathrm L}{\mathrm i}}

\newcommand{\eff}{{\mathrm e}{\mathrm f}{\mathrm f}}

\usepackage{graphicx}
\usepackage{dcolumn}
\usepackage{bm}

\begin{document}

\title{Neutron Diffusion and Nucleosynthesis in an Inhomogeneous Big Bang 
       Model}

\author{Juan F. Lara}
\email{ljuan@clemson.edu}
\affiliation{Department of Physics and Astronomy, Clemson University \\
             Clemson, South Carolina 29631}

\date{\today}

\begin{abstract}
This article presents an original code for Big Bang Nucleosynthesis in a baryon
inhomogeneous model of the universe.  In this code neutron diffusion between 
high and low baryon density regions is calculated simultaneously with
the nuclear reactions and weak decays that compose the nucleosynthesis 
process.  The size of the model determines the time when neutron diffusion 
becomes significant.  This article describes in detail how the time of neutron
diffusion relative to the time of nucleosynthesis affects the final abundances
of $ ^{4}$He, deuterium and $ ^{7}$Li.  These results will be compared with the
most recent observational constraints of $ ^{4}$He, deuterium and $ ^{7}$Li. 
This inhomogeneous model has $ ^{4}$He and deuterium constraints in concordance
for baryon to photon ratio $\eta = ( 4.3 - 12.3 ) \times 10^{-10}$  $ ^{7}$Li
constraints are brought into concordance with the other isotope constraints by
including a depletion factor as high as 5.9.  These ranges for the baryon to 
photon ratio and for the depletion factor are larger than the ranges from a 
Standard Big Bang Nucleosynthesis model.
\end{abstract}



\maketitle

\section{Introduction}

Big Bang Nucleosynthesis ( BBN ) is the primary mechanism of the creation of 
the lightest isotope species \cite{skm93,steigman03a}.  At temperatures of the 
universe $T \approx $ 100 GK baryonic matter consisted mostly of free neutrons
and protons in thermal equilibrium with each other via weak interconversion
reactions.  Weak freeze-out occurs when the temperature falls to $T \approx $ 
13 GK and the interconversion reactions fall out of equilibrium.  Between 
$T \approx $ 13 GK and $T \approx $ 0.9 GK only neutron decay changes the 
neutron and proton abundances.  Then nuclear reactions become significant, 
forming heavier and heavier nuclei.  Nearly all free neutrons at the time of 
nucleosynthesis are incorporated into $ ^{4}$He nuclei because of the large 
binding energy of that nuclei.  The amount of free neutrons at that time 
depends on the neutron lifetime $\tau_{\neu}$.  BBN is also the only source of 
deuterium production, and a significant source of $ ^{7}$Li production.  

The nuclear reaction rates depend on the baryon energy density $\rho_{b}$, 
equivalently the baryon to photon ratio $\eta$.  The abundance results of 
$ ^{4}$He, deuterium and $ ^{7}$Li can then be compared with measurements to 
put observational constraints on the value of $\eta$.  BBN constraints on 
$\eta$ can be compared with constraints derived from Cosmic Microwave 
Background measurements 
\cite{lee01,jaffe01,netterfield02,halverson02,melchiorri02,bennett03}
Acoustic oscillations in the CMB angular power spectrum are fitted with 
spherical harmonic functions that depend on several cosmological parameters, 
including the density factor $\Omega_{b} h^{2}$.  The most recent CMB 
measurements set $\Omega_{B} h^{2} = 0.0224 \pm 0.0009$ \cite{bennett03}, 
corresponding to $\eta = ( 5.9 - 6.4 ) \times 10^{-10}$. 

The Standard Big Bang Nucleosynthesis ( SBBN ) model is the simplest BBN model.
In SBBN all constituents are homogeneously and isotropically distributed.  
Parameters that define the SBBN model are $\eta$, $\tau_{\neu}$, and the number
of neutrino species $N_{\nu}$.  But a variety of models alternative to SBBN can
be fashioned by adding in other parameters.  The ability for BBN to constrain
the value of $\eta$ depends on the reliability of isotope measurements.  
Isotope observational constraints on $\eta$ have frequently appeared not to be
in concordance with each other when applied to the SBBN model.  The possibility
of alternative BBN models resolving discrepencies has then been considered
\cite{stkgt94,hsstwbl95,cst95,ks96,hsbl97,ko98,kks99,emmp00,pagel00,jr01,ks01,
os01,emmmp01,kajino02,nl02,iykom02,bgsw02}.  A good understanding of 
alternative models should then be maintained.  

\begin{figure}
   \includegraphics[scale=0.8,angle=0]{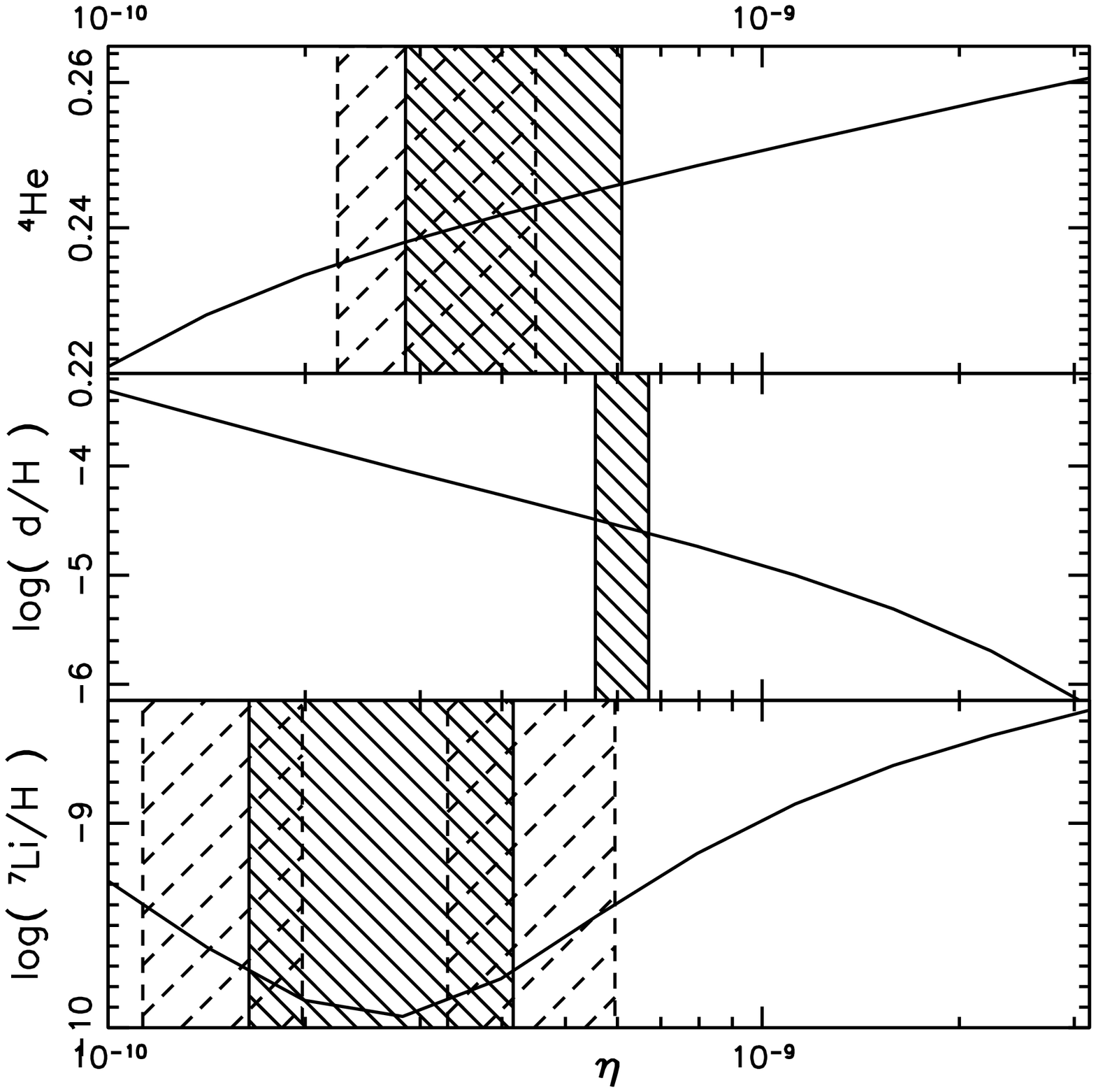}
   \caption{\label{fig:sbbnoc}
            $X_{ ^{4}\he}$, $Y(\deu)/Y(\pro)$ and $Y( ^{7}\li)/Y(\pro)$ for the
            SBBN case.  The graph for $X_{ ^{4}\he}$ includes the measurements 
            by IT04 ( \cite{it04} solid lines ) and by Luridinia et al 
            ( \cite{lppc03} dashed lines ).  The graph for 
            $Y( ^{7}\li)/Y(\pro)$ shows measurements by both Ryan et al 
            ( \cite{rbofn00} solid lines ) and Melendez \& Ramirez 
            ( \cite{mr04} dashed lines ).}
\end{figure}

Figure~\ref{fig:sbbnoc} shows graphs for the mass 
fraction $X_{ ^{4}\he}$ of $ ^{4}$He and the abundance ratios $Y(\deu)/Y(\pro)$
and $Y( ^{7}\li)/Y(\pro)$ of deuterium and $ ^{7}$Li, all as functions of 
$\eta$.  These graphs correspond to an SBBN model.  The SBBN code used for 
Figure~\ref{fig:sbbnoc} has been used by this author in previous articles 
\cite{lara98}.    

$ ^{4}$He is measured in low metallicity extragalactic HII regions.  There is
disagreement over how to extrapolate data points to zero metallicity.  In some
studies extrapolations have led to a higher mass fraction value of around 0.244
\cite{itl94,itl97,it98}, while in other studies the value is a lower 0.234
\cite{os95,oss97,ppr00}.   The most recent measurements of $X_{ ^{4}\he}$ have 
been a lower $0.239 \pm 0.002$ \cite{lppc03} and a higher $0.242 \pm 0.002$ 
\cite{it04}.  But the extent of systematic errors in these results is 
controversial.  Olive et al \cite{osw00} have used a compromise value 
$0.238 \pm 0.005$ combining both high and low measurements due the uncertainty 
in systematic error.  Recently Olive and Skillman \cite{os04} try to quantify 
uncertainties due to systematic error, reporting a large range 0.232 
$\le X_{ ^{4}\he} \le$ 0.258.  This range should eventually go down as the 
quantification of systematic errors improves.  Figure~\ref{fig:sbbnoc} shows 
the $2\sigma$ ranges by Luridinia et al \cite{lppc03} and Izotov and Thuan 
\cite{it04} (IT04) combined, corresponding to a range of 
$\eta = ( 2.2 - 6.1 ) \times 10^{-10}$

The deuterium measurement shown in Figure~\ref{fig:sbbnoc} is the weighted mean
of five Quasi-Stellar Objects ( QSO )'s done by Kirkman et al \cite{ktsol03}.
This abundance ratio $Y(\deu)/Y(\pro) = 2.78_{-0.38}^{+0.44} \times 10^{-5}$ is
in good agreement with many previous measurements 
\cite{tfb96,bt98a,bt98b,steigman01,otksplw01}.  But Rugers and Hogan 
\cite{rh96} measured $Y(\deu)/Y(\pro)$ an order of magnitude greater, at 
$( 1.9 \pm 0.4 ) \times 10^{-4}$. The abundance ratio by Kirkman et al 
corresponds to $\eta = ( 5.6 - 6.7 ) \times 10^{-10}$, which is in good 
agreement with the CMB results.

Ryan et al \cite{rbofn00} measure $ ^{7}$Li by looking at a group of very 
metal-poor stars and accounting for various systematic errors to derive a value
$Y( ^{7}\li)/Y(\pro) = 1.23_{-0.32}^{+0.68} \times 10^{-10}$.  This measurement
has a smaller magnitude and value of $\sigma$ than preceding measurements
\cite{pwsn99,pswn02}.  The largest uncertainty in the calculation of this 
abundance range is the uncertainty in determining the effective temperature
$T_{\eff}$ of the stars.  Melendez \& Ramirez \cite{mr04} make new calculations
of $T_{\eff}$ and get higher temperatures than Ryan et al for lower metallicity
stars.  Melendez \& Ramirez then derive a larger value 
$Y( ^{7}\li)/Y(\pro) = 2.34_{-0.96}^{+1.64} \times 10^{-10}$, also with a 
larger $2\sigma$ error.  Figure~\ref{fig:sbbnoc} shows the measurements by 
both Ryan et al and Melendez \& Ramirez.  Ryan et al's measurement corresponds 
to $\eta = ( 1.6 - 4.2 ) \times 10^{-10}$ while Melendez \& Ramirez's 
measurement can correspond to two ranges, 
$\eta = ( 1.1 - 2.0 ) \times 10^{-10}$ and 
$\eta = ( 3.3 - 6.0 ) \times 10^{-10}$.

This measurement of $ ^{4}$He by IT04 is in concordance with the deuterium 
measurement of Kirkman et al only at its $2\sigma$ range, for a narrow range
$\eta = ( 5.6 - 6.1 ) \times 10^{-10}$.  The $ ^{7}$Li constraints by Melendez
\& Ramirez is in concordance with the deuterium measurement also only at its 
$2\sigma$ range, while the $ ^{7}$Li constraints by Ryan et al have no region
of concordance at all.  A depletion factor from stellar evolution could improve
concordance between the deuterium constraints and the $ ^{7}$Li constraints by 
Melendez \& Ramirez.  A factor of 2.8 would resolve the discrepency in the case
of Ryan et al's constraints.  But models for $ ^{7}$Li depletion in stars and 
measurements of a depletion factor remain controversial.

This article focuses on the particular alternative model of Big Bang 
Nucleosynthesis with an Inhomogeneous baryon distribution ( IBBN ).  The IBBN
code used in this article is an original code written by this author 
\cite{lara01a}, hereafter known as the Texas IBBN code.  Upon publication of
this article the code will be made publically available at the author's website
\cite{website}.  The Texas IBBN code can serve as a consistency check against 
other IBBN codes, and against SBBN codes as well when run in its small distance
scale limit.  

Section~\ref{sec:hist} is a summary of the history of IBBN research, 
emphasizing developments that are significant to the way the Texas IBBN code is
constructed.  Section~\ref{sec:det} lists the specific details of the IBBN 
model used for this article.  Section~\ref{sec:res} shows the final abundance
results of the IBBN code for a range of distance scale $r_{i}$ and baryon to
photon ratio $\eta$.  This section discusses how the time of neutron diffusion
relative to weak freeze-out and nucleosynthesis significantly affects the final
isotope abundances the code produces.  The description of this relation in this
article is a useful guide for how baryonic matter flows and is processed in an
IBBN model.  In Section~\ref{sec:oc} the IBBN model will be compared with the
most recent constraints on $ ^{4}$He, deuterium and $ ^{7}$Li.  For certain 
IBBN parameter values the acceptable range of $\eta$ from $ ^{4}$He and 
deuterium constraints is widened.  The IBBN model also permits a large range of
$ ^{7}$Li depletion factor that is of particular interest.

\section{Development of the IBBN Code}

\label{sec:hist}

Various theories of baryogenesis lead to inhomogeneous distributions of free
neutrons and protons by the time of nucleosynthesis.  The distributions can be
modelled with many different symmetries.  Baryon inhomogeneities may arise from
a first order quark hadron phase transition \cite{witten84,kurkisuonio88,is01}.
Transport of baryon number between quark gluon phase and hadronic phase is 
inefficient, leading to concentration of baryon number in the last remaining 
regions regions of quark gluon plasma \cite{mm93}.  The magnitude of the bubble
surface tension determines if the quark gluon plasma regions form in centrallly
condensed spherical bubbles \cite{tsommf94,kks99} or cylindrical filaments 
\cite{okbm97}.  A cosmic string moving through matter during the quark hadron 
phase transition can also leave wakes of matter that remain in the quark gluon
plasma phase longer than in the regions outside the wakes, forming sheets of 
planar symmetric inhomogeneity \cite{lss01,lss03}.  Baryon inhomogeneity may 
also form during the earlier electroweak phase transition 
\cite{fjmo94,heckler95,kks99}.  A first order phase transition would proceed 
by bubble nucleation.  Particles in the plasma interact with the bubble walls 
in a CP violating matter, leading to a baryon asymmetry forming along the 
walls \cite{rs96,cjk98}.  The baryon density is in the form of high density 
shells, spherical or cylindrical \cite{okbm97,kks99}.  Baryon inhomogeneities
can also arise from phase transitions involving inflation-generated 
isocurvature fluctuations \cite{ds93}, or kaon condensation phase 
\cite{nelson90}.

The earliest articles on inhomogeneous codes \cite{zeldovich75,ep75,bm83} 
treated regions of different density as separate SBBN models.  They would run a
model with a high value of $\rho_{b}$, then a model with a low value, and then 
average the mass fractions from each model together, weighting each on how 
large each density region was.  Applegate, Hogan and Scherrer \cite{ahs87} 
considered the possibility of nucleons diffusing from high density regions to 
low density regions.  Neutrons diffuse by scattering off of electrons and 
protons.  Protons scatter off of neutrons and Coulomb scatter off of electrons,
but the mean free path of protons is about $10^{6}$ times smaller than that for
neutrons because of the Coulomb scattering.  Diffusion of other isotopes is 
negligible compared to neutron scattering because the isotopes are much more 
massive. 

In early IBBN codes that featured neutron diffusion \cite{ahs87,afm87,kb90} 
the diffusion part is run first, at early times and high temperatures.  Then 
nucleosynthesis within the regions is allowed to run.  In their IBBN code 
Kurkio-Suonio et al \cite{kmcrw88} (KMCRW88) made the significant innovation of
having neutron diffusion occur both before and during nucleosynthesis.  This
code was for planar symmetric baryon inhomogeneity, and was split in a uniform 
grid of 20 zones.  Kurki-Suonio and Matzner \cite{km89} (KM89) and Kurki-Suonio
et al \cite{kmos90} (KMOS90) looked at cylindrical and spherical models, using
uniform grids as well.  But for larger ratios between high and low densities, 
or lower volume fractions of high density region, the number of zones needed 
for the code to run accurately increased considerably.  The codes used by 
Kurki-Suonio and Matzner \cite{km90} (KM90) and Mathews et al 
\cite{mmaf90,mko96} instead use nonuniform grids, with a greater number of 
narrower zones around the boundary between high and low density regions, where
they are needed.  Mathews et al \cite{mmaf90} halve the width of a zone the 
closer the zone is to the boundary.  KM90 use a stretching function to make a 
grid of 64 zones that get very narrow around the boundary.

\section{The IBBN Model}

\label{sec:det}

In an IBBN model the universe is represented as a lattice of baryon 
inhomogeneous regions.  An IBBN code models one region in that lattice.  The 
inhomogeneity can have planar, cylindrical or spherical symmetry.  The Texas
IBBN code has been used to model condensed spheres \cite{lara01b} and cylinders
( a high density core ) and spherical and cylindrical \cite{lara04} shells ( a 
high density outer layer ).  The parameters that define an IBBN model are the 
baryon to photon ratio $\eta$, the distance scale $r_{i}$, the density contrast
$R_{\rho}$, and the volume fraction $f_{v}$.  The distance scale is the 
initial size of the model at a chosen time.  In this article that time is the 
starting time of a run, when the temperature $T = $ 100 GK.  The density 
contrast is the initial ratio of high baryon density to low density.  The 
volume fraction is the fraction of the model occupied by the high density 
region.  $f_{v}$ is parametrized such that it corresponds to a specific radius.

A cylindrical shell model will be used in this article.  This symmetry has been
used by Orito et al \cite{okbm97} and Lara 2004 \cite{lara04}.  The isotope 
abundance results are represented as contour maps in a parameter space defined 
by $\eta$ and $r_{i}$.  The values of the remaining parameters are taken from 
Orito et al \cite{okbm97}.  

\begin{eqnarray*}
               R_{\rho} & = & 10^{6} \\
   1 - \sqrt{1 - f_{v}} & = & 0.075
\end{eqnarray*}

\noindent The contour lines of abundance values to be discussed in 
Sections~\ref{sec:res} and \ref{sec:oc} are most greatly exaggerated in a 
cylindrical shell model with the parameter values from Orito et al, meaning 
that observational constraints will be satisfied for the highest possible 
values of $\Omega_{B} h^{2}$ ( $\eta$ ).  The parametrization of $f_{v}$ means 
that the thickness of the high density outer shell equals 0.075 the radius of 
the whole model.  For the neutron lifetime the most recent world average 
$\tau_{\neu} = $ 885.7 seconds \cite{eidelman04} is used. 

The model is divided into a core and 63 cylindrical shells.  These zones need 
to be thin at the boundary radius $r_{b}$ between high and low density to 
accurately model neutron diffusion.  The Texas IBBN code uses the stretching
function from KM90 \cite{km90} to set the radii of the shells.

\begin{eqnarray}
   \xi(r) & = & \xi(r_{b}) + \frac{1}{C_{1}} 
                \left ( 1 - \frac{1}{C_{3}} \right )
                \sqrt{\frac{C_{2}}{C_{3}}} \arctan
                \left [ ( r - r_{b} ) \sqrt{\frac{C_{3}}{C_{2}}} \right ] +
                \frac{r - r_{b}}{C_{1} C_{2}}
\end{eqnarray}

\begin{figure}
   \includegraphics[scale=0.9,angle=90]{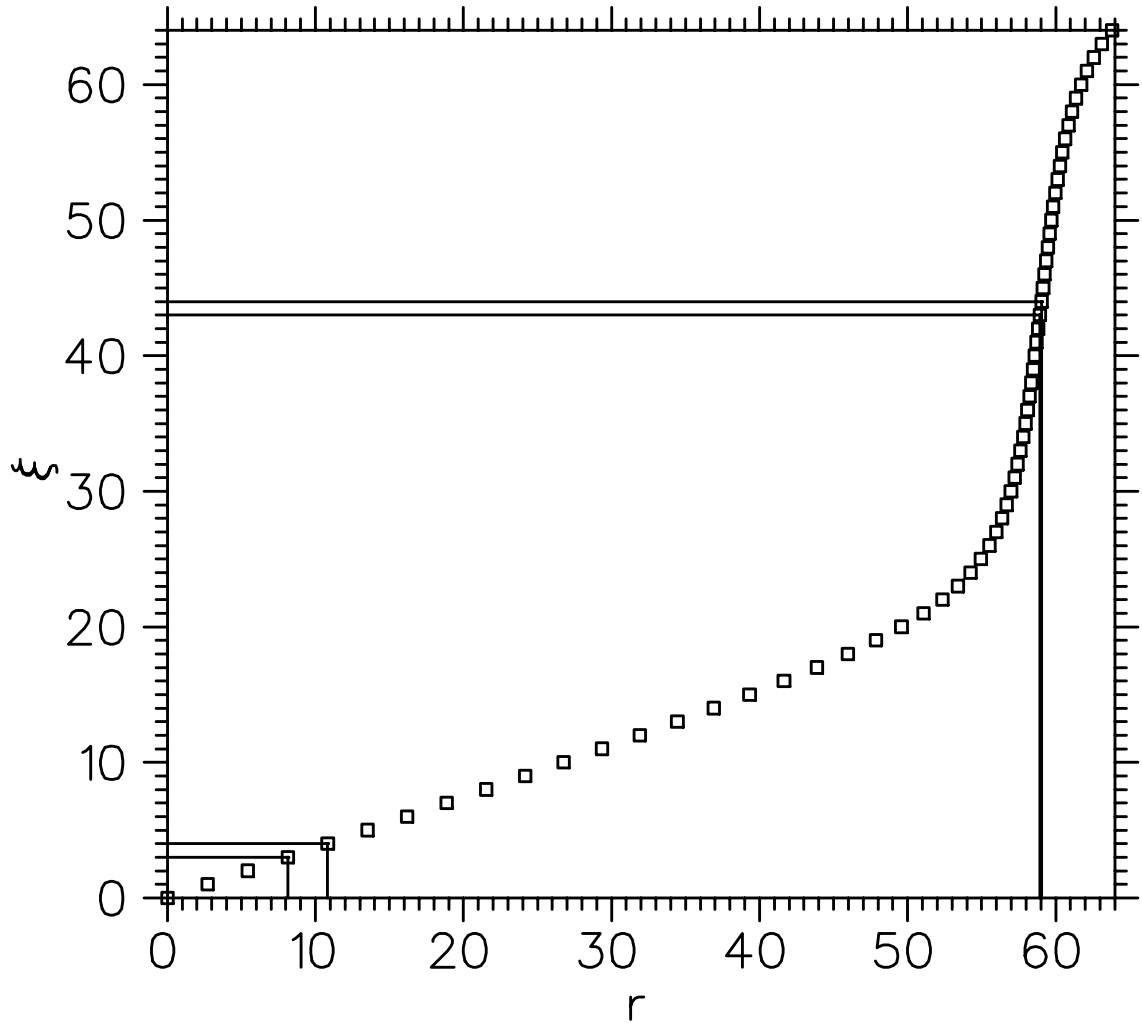}   
   \caption{\label{fig:grid} The stretching function will map integer values 
            of $\xi$ to values of radius $r$ such that the $r$'s near the 
            boundary radius $r_{b}$. are closely spaced.  Here $r_{b} = $ 
            59.2.}
\end{figure}

\noindent $\xi(r)$ is the shell number out from the center, with a radius $r$ 
in normalized units that range from 0 to 64.  The boundary radius $r_{b}$ = 
59.2 as determined by the value of $f_{v}$  Figure~\ref{fig:grid} shows how
 $\xi(r)$ maps onto $r$

Appendix~\ref{app:trace} describes in detail the method the calculations are 
made for each timestep in the run.

\section{Results}

\label{sec:res}

Figures~\ref{fig:cmhe4}-\ref{fig:cmli7} are contour maps of the overall mass 
fraction $X_{ ^{4}{\he}}$ and abundance ratios $Y(\deu)/Y(\pro)$ and 
$Y( ^{7}\li)/Y(\pro)$ at the end of the Texas IBBN code's run, drawn in a 
parameter space defined by $\eta$ and $r_{i}$.  In Figure~\ref{fig:cmli7} the
abundance ratios of both $ ^{7}$Li and $ ^{7}$Be are shown combined, as all 
the $ ^{7}$Be has decayed to $ ^{7}$Li by now.  Neutron diffusion starts at the
boundary between the high density outer region and the low density inner 
region, and then progresses outwards to the outermost shell and inwards to the 
core.  The time neutron diffusion takes to homogenize neutrons determines the
shapes of the contour lines shown in Figures~\ref{fig:cmhe4}-\ref{fig:cmli7}.  
The two milestone times in element synthesis are the times of weak freeze-out 
and nucleosynthesis.  The contour lines can be described in terms of whether 
neutron diffusion occurs before weak freeze-out, between weak freeze-out and 
nucleosynthesis, or after nucleosynthesis.

\subsection{Before Weak Freeze-Out}

For the smallest distance scales $r_{i}$ neutron diffusion homogenizes neutrons
very early in a run.  Protons are still coupled with neutrons via the 
interconversion reactions.  In the high density outers shells these 
interconversion reactions run in the direction of converting protons to 
neutrons, to keep up with neutron diffusion.  The protons converted to neutrons
diffuse to the low density inner shells, where the the interconversion 
reactions run in the opposite direction, converting neutrons to protons.  
Protons are then homogenized along with neutrons, and the final abundances are
the same as the abundances from an SBBN model.

For larger $r_{i}$ neutron diffusion takes longer to affect all shells of the 
model.  At a distance scale of around 1600 cm the time when diffusion ends 
coincides with weak freeze-out.  Protons are not as coupled with neutrons as 
with smaller distance scales, and so are not completely homogenized by the time
when neutrons have been homogenized.  A larger proton density makes 
nucleosynthesis occur earlier in the outer shells.  For a given value of 
$\eta$ then the final abundance results are the results from an SBBN model with
earlier nucleosynthesis:  greater $ ^{4}$He, lesser deuterium, and greater 
$ ^{7}$Li and $ ^{7}$Be production.  That corresponds to the shift in the 
contour lines of Figures~\ref{fig:cmhe4}-\ref{fig:cmli7} to lower $\eta$ for 
distance scales $r_{i}$ from 1600 cm to 25000 cm.

\subsection{Between Weak Freeze-Out and Nucleosynthesis:  $ ^{7}$Li and
            $ ^{7}$Be}

In models with $r_{i}$ from $\approx$ 25000 cm to 3.2 $\times 10^{5}$ cm 
neutrons are homogenized at a time in between weak freeze-out and 
nucleosynthesis.  

If $r_{i} \approx $ 25000 cm, neutron diffusion becomes significant everywhere 
right around the time of weak freeze-out. 
Figures~\ref{fig:25e4diff}-\ref{fig:25e4back} show an example of how the 
various reactions in the code interact with one another.
Figures~\ref{fig:25e4diff}-\ref{fig:25e4back} correspond to shell number 62, a
high density outer shell that is two shells away from the outer edge of the 
model.  The reaction rates are normalized to the average baryon number density 
$n_{b0}$ and the expansion rate of the universe $\dot{\alpha_{R}}$.  

In Figure~\ref{fig:25e4diff} the neutron diffusion rates peak at around $T = $ 
10.0 GK and remain large up to $T = $ 3.0 GK.  The diffusion rate from shell 62
out to shell 61 is larger than the rate from shell 63 into shell 62 all through
that time.  The net effect is outflow of neutrons from shell 62, as it is 
happening in all high density shells at this time.  The peak temperature $T = $
10.0 GK is just after the temperature $T \approx $ 13 GK of weak freeze-out.  
Figure~\ref{fig:25e4diff} shows the rates for the reactions that convert 
neutrons to protons ( n $\rightarrow$ p ) and the rates that convert protons to
neutrons ( p $\rightarrow$ n ) in short dashed lines.  These rates are the same
as they would be in the SBBN model.  So no proton redistribution via these 
reactions is possible, and the proton number density in shell 62 remains high.

Figure~\ref{fig:25e4back} shows the nuclear reaction rate n + p 
$\leftrightarrow$ d + $\gamma$ in short dashed lines.  This reaction falls out 
of Nuclear Statistical Equilibrium ( NSE ) at $T = $ 0.9 GK, starting off the
chain of nucleosynthesis.  Because the proton number density in the outer 
shells is high the nuclear reactions go at faster rates than they would in 
the SBBN model.  Nucleosynthesis then occurs slightly earlier in the outer 
shells, depleting neutrons there.  This deficit of neutrons leads to back 
diffusion.  Figure~\ref{fig:25e4back} shows the rates of diffusion from shell
61 into shell 62, and from shell 62 out to shell 61.  The net effect is now a
concentration of neutrons in the high density shells.  Nearly all 
nucleosynthesis is concentrated in the outer shells.

$ ^{7}$Li is created primarily by the nuclear reactions t + $ ^{4}$He 
$\leftrightarrow ^{7}$Li + $\gamma$ and n + $ ^{7}$Be $\leftrightarrow$ p + 
$ ^{7}$Li, and destroyed primarily by the reactions p + $ ^{7}$Li 
$\leftrightarrow$ 2( $ ^{4}$He ) and d + $ ^{7}$Li $\leftrightarrow$ n + 
2( $ ^{4}$He ).  The depletion reaction p + $ ^{7}$Li $\leftrightarrow$ 
2( $ ^{4}$He ) dominates over other reactions involving $ ^{7}$Li.  $ ^{7}$Be 
is created primarily by $ ^{3}$He + $ ^{4}$He $\leftrightarrow ^{7}$Be + 
$\gamma$ and destroyed primarily by n + $ ^{7}$Be $\leftrightarrow$ p + 
$ ^{7}$Li.  In contrast to $ ^{7}$Li the creation reaction of $ ^{7}$Be 
dominates over the destruction reaction, and greater $ ^{4}$He production in 
the high density shells magnifies the dominance even further.
Figure~\ref{fig:Li7Be725E4} shows the number densities of $ ^{7}$Li and 
$ ^{7}$Be as functions of radius.  The number density of  $ ^{7}$Be is 
considerably larger in the high density outer shells than in the rest of the 
model.  Due to this greater $ ^{7}$Be production the contour lines in 
Figure~\ref{fig:cmli7} have a larger shift to lower $\eta$ than the contour 
lines in Figures~\ref{fig:cmhe4}-\ref{fig:cmdeu}.

\subsection{Between Weak Freeze-Out and Nucleosynthesis:  $ ^{4}$He and 
            deuterium}

\begin{figure}
   \includegraphics[scale=0.8,angle=90]{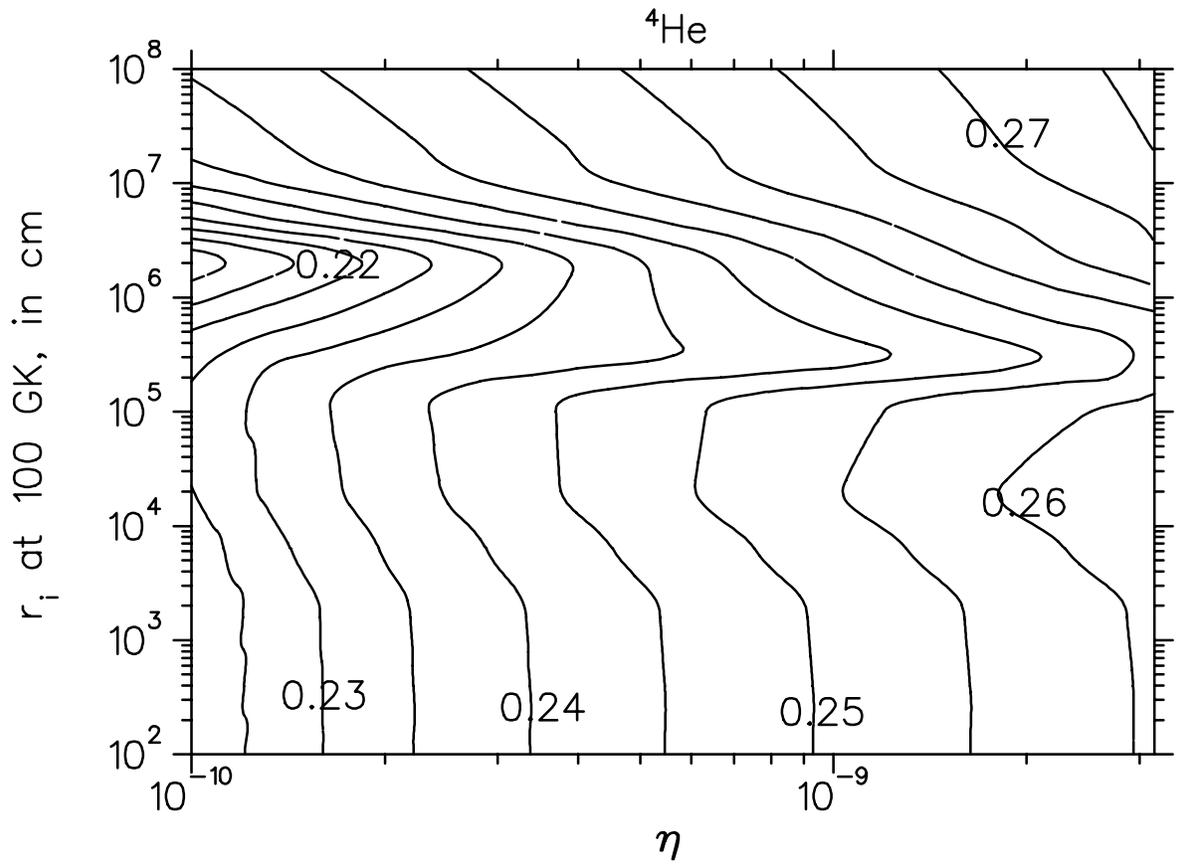}   
   \caption{\label{fig:cmhe4} The mass fraction $X_{ ^{4}\he}$ in the IBBN 
            code.  The horizontal axis is for baryon-to-photon ratio $\eta$ and
            the vertical axis is for distance scale $r_{i}$ in centimeters at 
            temperature $T = $ 100 GK.}
\end{figure}

In models with $r_{i}$ from $\approx$ 25000 cm to $10^{5}$ cm the proton 
number density is unchanged from the time of weak freeze-out to 
nucleosynthesis, except for a slight increase due to neutron decay.  For this 
range of $r_{i}$ the contour lines in Figures~\ref{fig:cmhe4}-\ref{fig:cmli7} 
lie along nearly constant values of $\eta$.  

In models with $r_{i} = 10^{5}$ cm the amount of time needed for back diffusion
to affect all shells is the same as the duration time of nucleosynthesis.  For
larger distance scales the shells furthest from the boundary are not as well 
coupled by back diffusion to the boundary shells.  Nucleosynthesis becomes 
concentrated in the shells immediately around the boundary.  This 
concentration leads to an overall drop in $ ^{4}$He production.  For $r_{i}$ 
from $\approx 10^{5}$ cm to 3.2 $\times 10^{5}$ cm the contour lines in 
Figures~\ref{fig:cmhe4}-\ref{fig:cmli7} shift to higher $\eta$.  
Figure~\ref{fig:FnlHe4} shows the final number density of $ ^{4}$He as a 
function of radius for $r_{i} \approx 3.2 \times 10^{5}$ cm, with  $ ^{4}$He
very concentrated around the boundary.  

For models $r_{i}> 3.2 \times 10^{5}$ cm diffusion cannot homogenize neutrons 
before nucleosynthesis.  A larger neutron number density remains in the 
outermost high density shells, and a lower density in the low density core and 
innermost shells.  The larger neutron number density leads to greater $ ^{4}$He
production in the outermost shells.  Figure~\ref{fig:FnlHe4} shows the final 
number density of $ ^{4}$He for $r_{i} = 2.0 \times 10^{6}$ cm.  There is 
greater $ ^{4}$He production around the boundary and the outermost shells and a
trough of lower production in between.  The overall $ ^{4}$He mass fraction 
increases again,  For $r_{i}> 3.2 \times 10^{5}$ cm the contour lines in 
Figure~\ref{fig:cmhe4} and Figure~\ref{fig:cmli7} shift to lower $\eta$.

Decreased $ ^{4}$He production tends to be accompanied by increased deuterium 
production.  Figure~\ref{fig:FnlDeu} shows the final number density of 
deuterium for $r_{i} \approx 3.2 \times 10^{5}$ cm and 
$r_{i} = 2.0 \times 10^{6}$ cm.  In the radii corresponding to the trough of
$ ^{4}$He production in Figure~\ref{fig:FnlHe4} Figure~\ref{fig:FnlDeu} has a 
peak in deuterium production.  The contour lines in Figure~\ref{fig:cmdeu} 
shift to higher $\eta$ for $r_{i}$ from $\approx 10^{5}$ cm to 3.2 
$\times 10^{5}$ cm, just as in Figure~\ref{fig:cmhe4} and 
Figure~\ref{fig:cmli7}.  But for $r_{i}$ from $\approx 3.2 \times 10^{5}$ cm to
2.0 $\times 10^{6}$ cm the deuterium contour lines still shift to higher $\eta$
because of the increased deuterium production shown in Figure~\ref{fig:FnlDeu}.

\subsection{After Nucleosynthesis}

At $r_{i} \approx 2.0 \times 10^{6}$ cm neutron diffusion peaks at the same 
time as nucleosynthesis.  For models with larger $r_{i}$ neutron diffusion 
becomes less significant.  More neutrons initially in the high density outer 
region remain there, increasing $ ^{4}$He production.  The trough in 
Figure~\ref{fig:FnlHe4} disappears and so deuterium production decreases.  In 
Figure~\ref{fig:cmdeu} the deuterium contour lines shift to lower $\eta$ to 
coincide with the contour shifts in Figure~\ref{fig:cmhe4} and 
Figure~\ref{fig:cmli7}.  The largest models behave as two separate SBBN models;
a high density SBBN model with considerable $ ^{4}$He and $ ^{7}$Li$+^{7}$Be 
production and minimal deuterium production, and a low density SBBN model with 
minimal $ ^{4}$He and $ ^{7}$Li$+^{7}$Be production and substantial deuterium 
production.  Final results are the average results from the two models.

\subsection{Generalization}

The contour maps shown in Figures~\ref{fig:cmhe4}-\ref{fig:cmli7} are for a 
specific IBBN model.  If the model geometry is changed or if the values of the 
other parameters, the density contrast $R_{\rho}$ and the volume fraction 
$f_{v}$, are changed the shifts in the contour lines become more or less 
exaggerated.  But the basic shapes of the contour lines persist.  For all 
geometries and values of $R_{\rho}$ and $f_{v}$ there will be a range of 
distance scale where neutron homogenization occurs in the interim between weak 
freeze-out and nucleosynthesis, leading to the shift to lower $\eta$ as shown 
in this article's model for $r_{i} \approx $ 25000 cm.  There will also be a 
range of $r_{i}$ where neutron diffusion coincides with nucleosynthesis.  A 
trough of lower  $^{4}$He production between the boundary and the high density 
shells furthest from the boundary develops in this range, like the trough shown
in Figure~\ref{fig:FnlHe4}.  IBBN models will then have a distance scale where 
the contour lines of $ ^{4}$He and deuterium diverge.  For a talk at the Sixth 
ResCEU International Symposium \cite{lara04} this author looked at models with 
the geometries of condensed cylinders, condensed spheres, and spherical shells 
as well as cylindrical shells.  The values of $R_{\rho}$ and $f_{v}$ used by 
Orito et al \cite{okbm97} were used in those runs.  The contour maps in all the
models showed the same features as seen in 
Figures~\ref{fig:cmhe4}-\ref{fig:cmli7}.

\section{Observational Constraints}

\label{sec:oc}

Figure~\ref{fig:ibbnocry} shows the observational constraints from 
Figure~\ref{fig:sbbnoc} applied to the contour maps of 
Figures~\ref{fig:cmhe4}-\ref{fig:cmli7}.  The maximum $X_{ ^{4}\he} \le $
0.246 constraint from IT04 \cite{it04} and the $ ^{7}$Li constraints from 
Ryan et al \cite{rbofn00} are shown in Figure~\ref{fig:ibbnocry}.

Regions of concordance between the IT04 $ ^{4}$He maximum constraint and the 
deuterium constraints are shown in yellow.  A concordance region exists for 
distance scales $r_{i} \le $ 5000 cm and 
$\eta = ( 5.6 - 6.1 ) \times 10^{-10}$.  These limits on $\eta$ are the same
limits as seen in SBBN models.  The maximum limit of $r_{i}$ is set by the 
shift to lower $\eta$ as neutron diffusion occurs closer to weak freeze-out.
The $X_{ ^{4}\he} = $ 0.246 contour have a greater shift than the contour lines
for the deuterium constraints, because of increased $ ^{4}$He production in the
outer region.  

Another region of concordance appears for $r_{i} = ( 1.3 - 6.0 ) \times 10^{5}$
cm, when the contour lines shift to higher $\eta$ due to the concentration of 
nucleosynthesis along the boundary.  The upper cutoff of $r_{i}$ is determined 
by the condition when a trough as shown in Figure~\ref{fig:FnlHe4} exists in
the $ ^{4}$He abundance distribution.  Greater $ ^{4}$He production in the 
outermost shells cause the $X_{ ^{4}\he} = $ 0.246 contour to shift to lower
$\eta$ while greater deuterium production in the trough cause the deuterium
contour lines to remain shifted to higher $\eta$.  The acceptable range of
$\eta$ is $( 4.3 - 12.0 ) \times 10^{-10}$, larger than in the SBBN case.

The $ ^{7}$Li constraints from Ryan et al \cite{rbofn00} are shown in darkest 
green in Figure~\ref{fig:ibbnocry}.  The contour lines for $ ^{7}$Li tend to 
shift in the same direction with the contour lines of $ ^{4}$He and deuterium.
So the $ ^{7}$Li constraints do not have a region of concordance with the 
$ ^{4}$He and deuterium constraints for this IBBN model, and the lack of a 
region of concordance persists for other geometries and parameter values.  
Figure~\ref{fig:ibbnocry} also shows the region of the $ ^{7}$Li constraints 
with a depletion factor of 2.8.  That depletion factor would bring the 
$ ^{7}$Li  constraints in concordance with the other isotopes for distances 
scales $r_{i} \le $ 5000 cm.  For the region of concordance corresponding to
$r_{i} = ( 1.3 - 6.0 ) \times 10^{5}$ cm a larger depletion factor of 5.9 is 
needed.  The greater production of $ ^{7}$Be shown in 
Figure~\ref{fig:Li7Be725E4} leads to the larger shift to lower $\eta$ in the 
$ ^{7}$Li contour lines compared to the $ ^{4}$He and deuterium contour lines,
and the larger depletion factor.  Figure~\ref{fig:ibbnocry} shows the region of
the $ ^{7}$Li constraints with the depletion factor of 5.9.  The benefit of 
IBBN models then is to allow for a larger range of $ ^{7}$Li depletion factor 
than permitted by the SBBN model.

Figure~\ref{fig:ibbnocry} is similar to Figure 2 from the proceedings article
Lara 2004 \cite{lara04}.  Differences between the figures include use of the 
newer $X_{ ^{4}\he} \le $ 0.246 constraint \cite{it04} in place of the 
$X_{ ^{4}\he} \le $ 0.248 constraint \cite{osw00}.  The method of calculating 
the diffusion coefficients \cite{jr01} is also newer than the method 
\cite{kagmbcs92} used in Lara 2004.  The neutron lifetime $\tau_{\neu} = $ 
885.7 seconds \cite{eidelman04} was also updated for this article.

Figure~\ref{fig:ibbnocmr} shows the $ ^{7}$Li constraints from Melendez \& 
Ramirez \cite{mr04}.  The outermost edge of these $2\sigma$ constraints has
concordance with most of the concordance region between $ ^{4}$He and deuterium
for $r_{i} \le $ 5000 cm.  With a small depletion factor of 1.35 these 
$ ^{7}$Li constraints cover the whole region of concordance.  A larger 
depletion factor of 2.8 is needed to cover the region of concordance 
corresponding to $r_{i} = ( 1.3 - 6.0 ) \times 10^{5}$ cm.  The regions of 
concordance between $ ^{4}$He and deuteurium are controversial because of 
considerable disagreement regarding $ ^{4}$He constraints.  Nonetheless both 
Figure~\ref{fig:ibbnocry} and Figure~\ref{fig:ibbnocmr} show that Inhomogeneous
Big Bang Nucleosynthesis allows for a larger range of acceptable $ ^{7}$Li 
depletion factor to bring deuterium and $ ^{7}$Li in concordance with each 
other, due to the greater shift in $ ^{7}$Li contour lines to lower $\eta$ for
distance scales $r_{i}$ from $\approx$ 1600 cm to $10^{5}$ cm.

\section{Conclusions}

The Texas IBBN code is an original code written such that the weak and nuclear 
reactions of element synthesis are coupled with neutron diffusion.  The time of
neutron diffusion relative to the times of weak freeze-out and nucleosynthesis
have a significant influence on the final production amounts of $ ^{4}$He, 
deuterium, and $ ^{7}$Li.  Because diffusion is coupled to the reaction network
the code correctly accounts for neutron back diffusion, wherein neutrons flow
back into regions with higher proton density due to earlier nucleosynthesis in
those regions.  Back diffusion has an influence over the results especially
when the time of neutron diffusion is close to the time of nucleosynthesis.  Of
most interest in the results is the larger range of depletion factor for 
$ ^{7}$Li that the IBBN model permits over the SBBN model.

In models where diffusion homogenizes the neutron distribution before weak 
freeze-out protons are coupled with the neutrons via the weak interconversion
reactions.  Protons are then redistributed.  Proton redistribution is less 
effective in models with the time of diffusion closer to the time of weak 
freeze-out, leaving a higher proton density in the outer shells.  Increasing 
proton density leads to earlier nucleosynthesis in the outer shells.  Neutrons 
then back diffuse into the outer shells, concentrating nucleosynthesis there.  
Nucleosynthesis in the high density shells produces decreasing amounts of 
deuterium and increasing amounts of $ ^{4}$He and especially $ ^{7}$Be.  The 
increased production of $ ^{7}$Be is significant in the determination of the 
depletion factor of $ ^{7}$Li.

\begin{figure}
   \includegraphics[scale=0.8,angle=90]{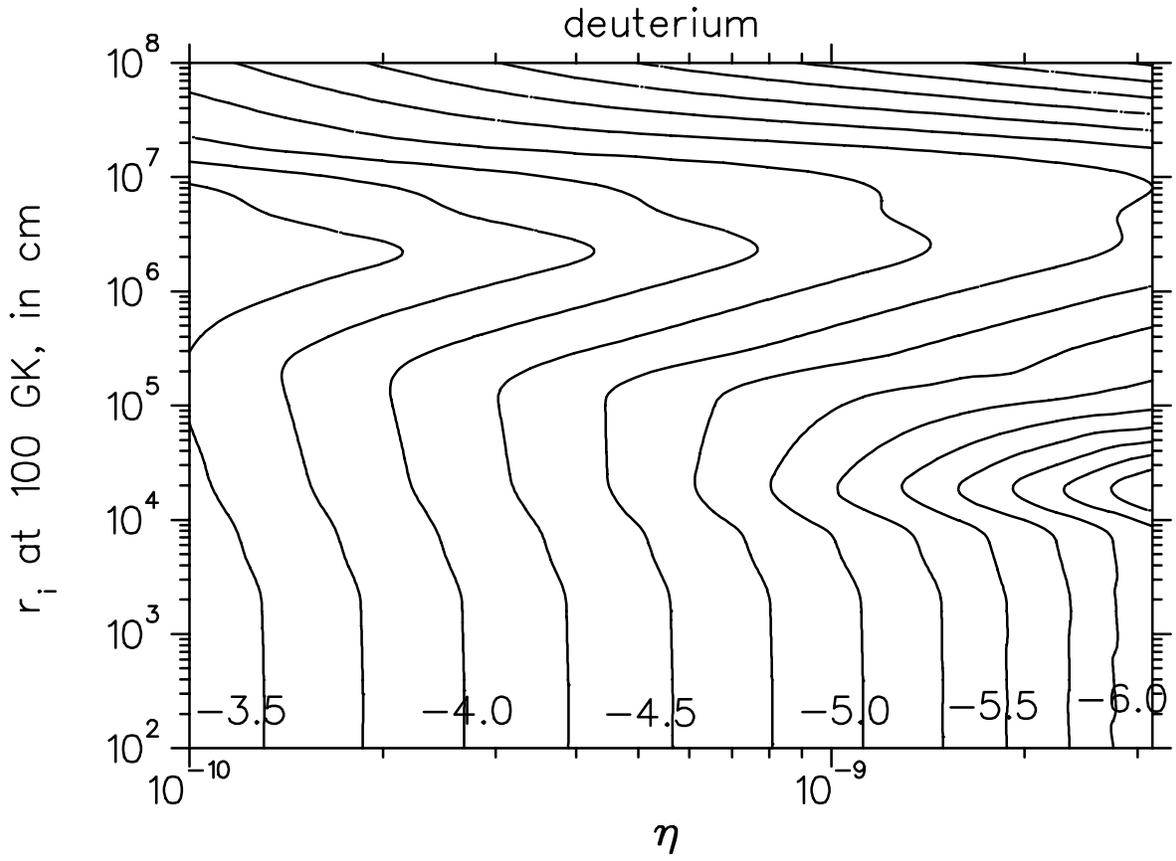}   
   \caption{\label{fig:cmdeu} The log of abundance $Y(\deu)/Y(\pro)$}
\end{figure}

\begin{figure}
   \includegraphics[scale=0.8,angle=90]{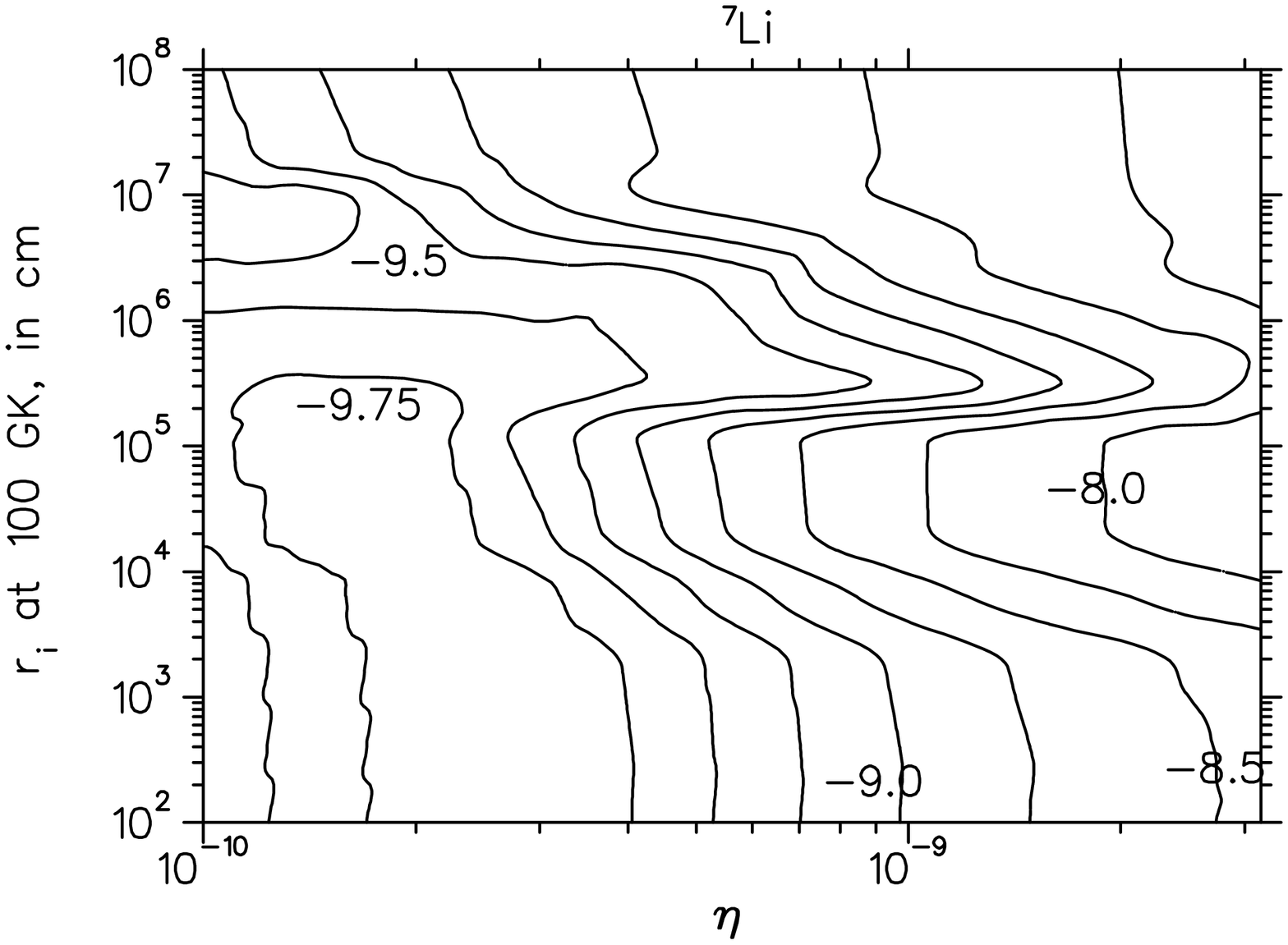}   
   \caption{\label{fig:cmli7} The log of abundance $Y( ^{7}\li)/Y(\pro)$}
\end{figure}

\begin{figure}
   \includegraphics[scale=0.8,angle=0]{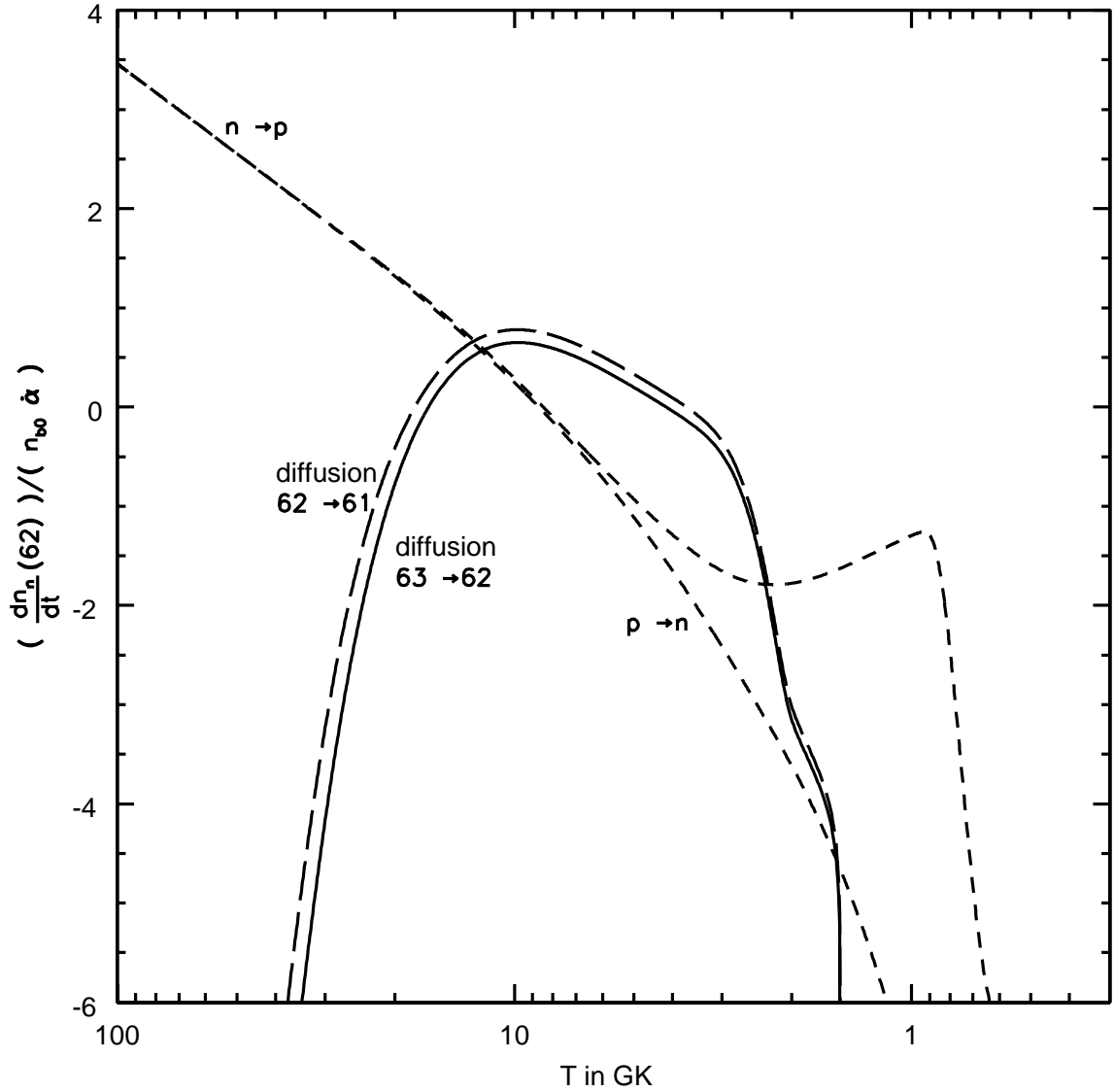}
   \caption{\label{fig:25e4diff} Neutron diffusion vs. the interconversion 
            reactions for $r_{i} \approx $ 25000 cm.  The rate of diffusion 
            from shell 63 to shell 62 is shown in solid while the greater rate 
            from shell 62 to shell 61 is shown in long dashed lines.  The net 
            result of diffusion is neutron depletion in shell 62.  The forward 
            and reverse direction rates of the neutron to proton conversion 
            reactions are shown in short dashed lines.  Neutrons and protons
            are decoupled during the peak time of diffusion.}
\end{figure}

\begin{figure}
   \includegraphics[scale=0.8,angle=0]{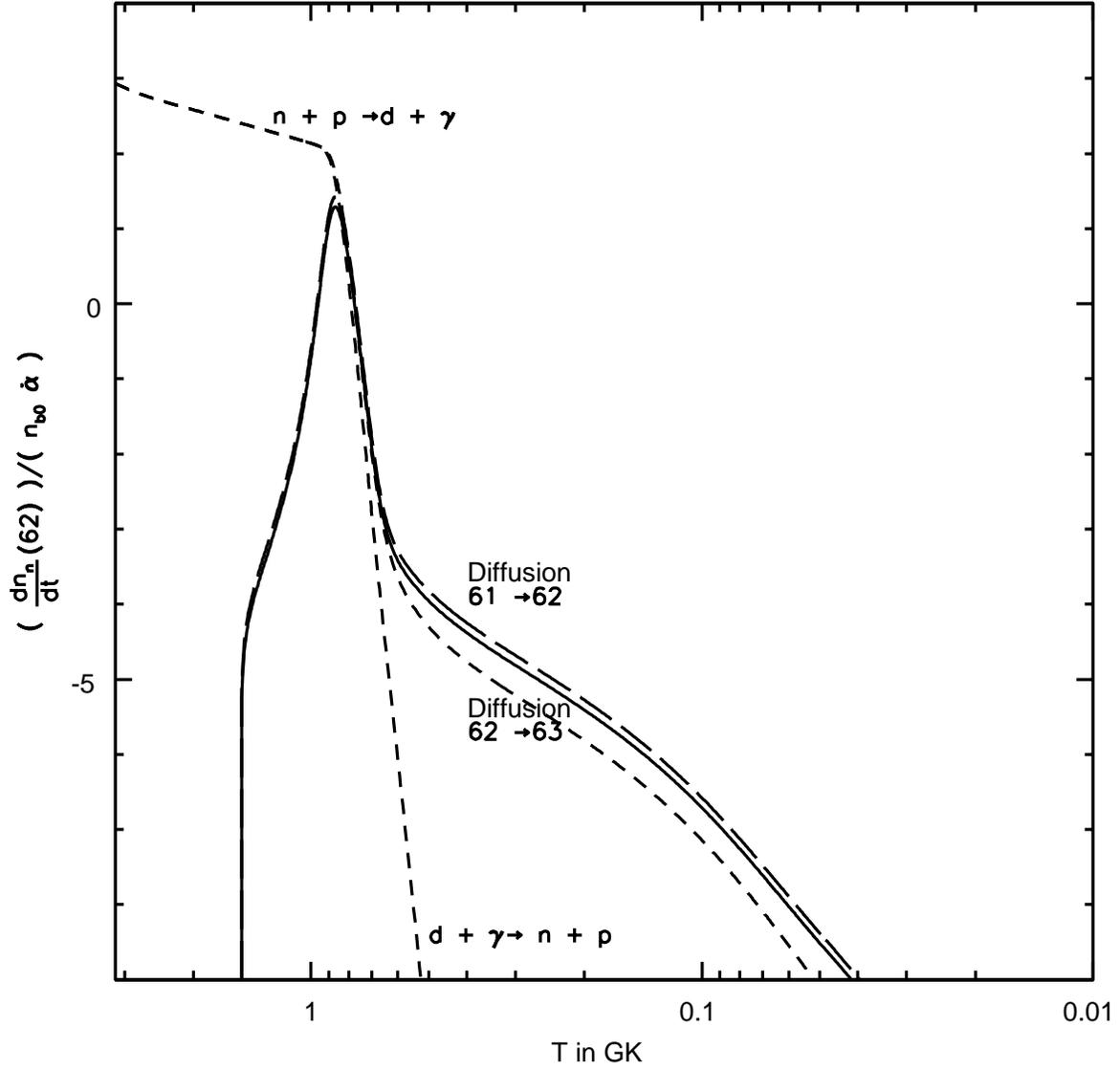}
   \caption{\label{fig:25e4back} Back diffusion for $r_{i} \approx $ 25000 cm.
            The rate of diffusion from shell 62 to shell 63 is shown in solid 
            while the greater rate from shell 61 to shell 62 is shown in long 
            dashed lines.  The direction of diffusion is now reversed, 
            resulting in a net increase of neutrons in shell 62.  Back 
            diffusion coincides with the time when the nuclear reaction n + p 
            $\leftrightarrow$ d + $\gamma$ ( shown in short dashed lines ) 
            falls out of NSE.}
\end{figure}

\begin{figure}
   \includegraphics[scale=0.8,angle=90]{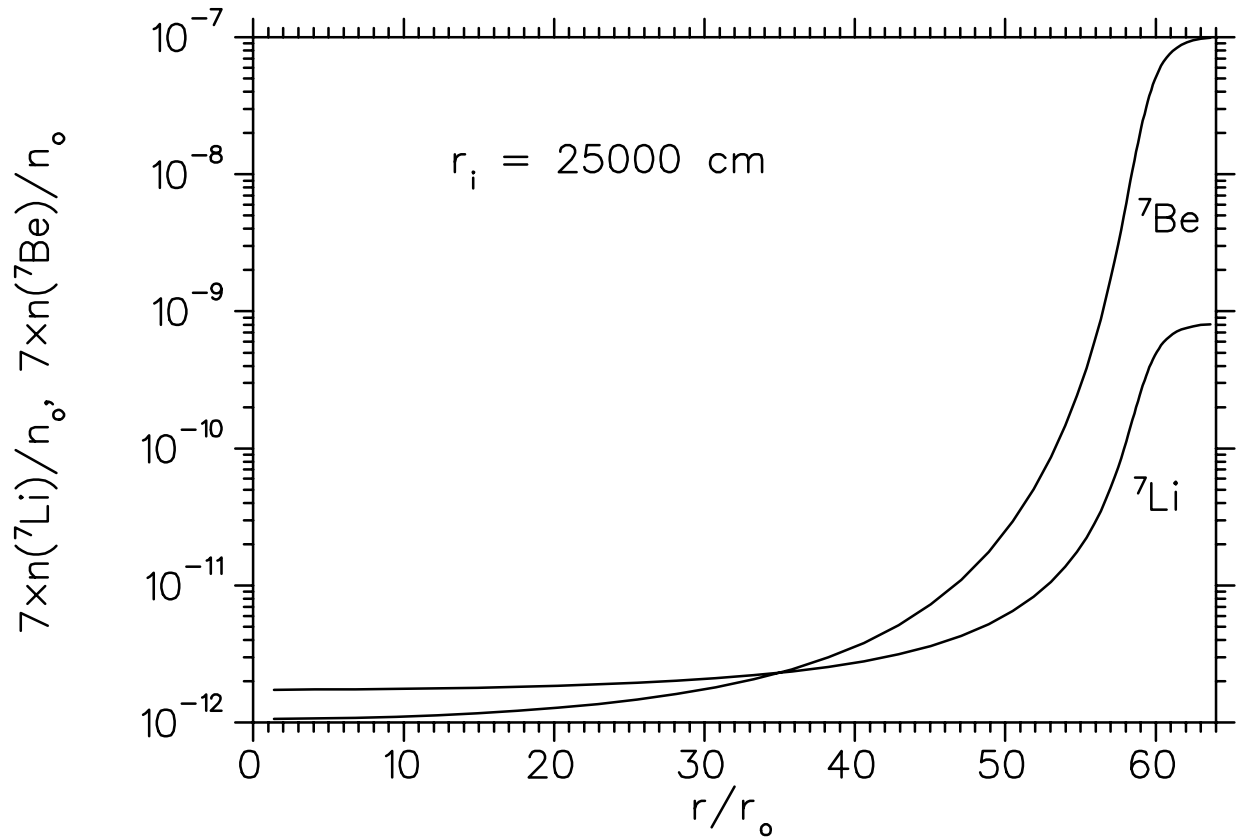}
   \caption{\label{fig:Li7Be725E4} The final number densities of $ ^{7}$Li and
            $ ^{7}$Be produced in each shell of the model, for $r_{i} \approx $
            25000 cm.}
\end{figure}

\begin{figure}
   \includegraphics[scale=0.8,angle=90]{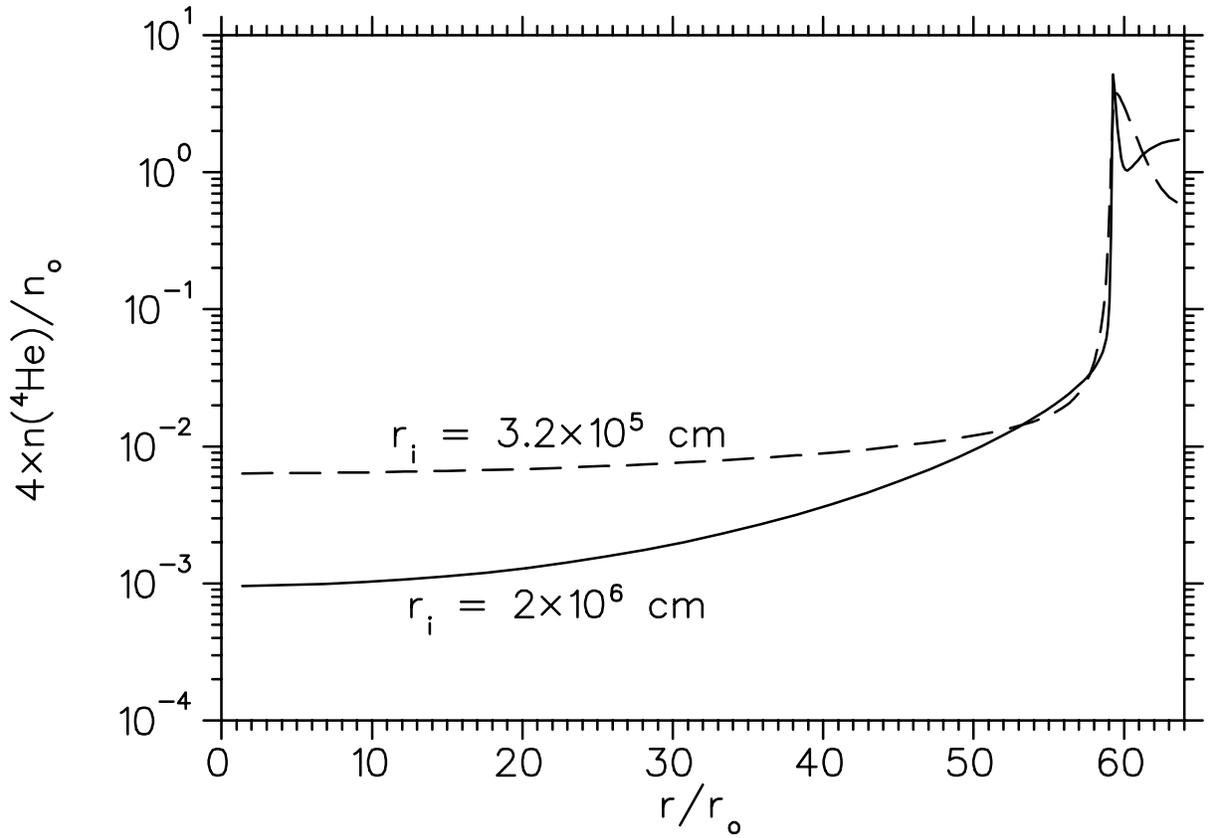}
   \caption{\label{fig:FnlHe4} The final number density of $ ^{4}$He produced 
            in each shell of the model.  For $r_{i} \approx 3.2 \times 10^{5}$
            cm $ ^{4}$He production is concentrated surrounding the boundary.
            For $r_{i} = 2 \times 10^{6}$ cm $ ^{4}$He production has incresed
            in the outermost shells due to neutrons not homogenizing.}
\end{figure}

\begin{figure}
   \includegraphics[scale=0.8,angle=90]{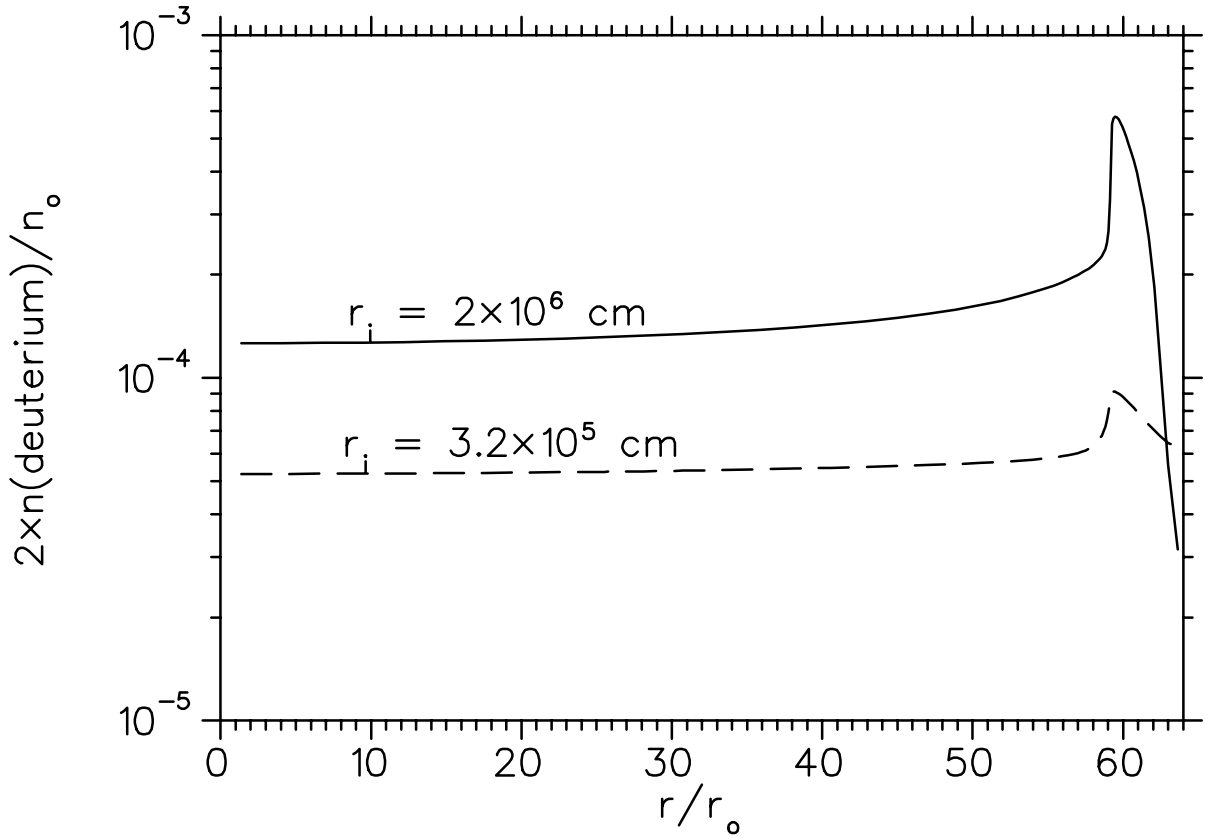}
   \caption{\label{fig:FnlDeu} The final number density of deuterium produced 
            in each shell of the model, for $r_{i} \approx 3.2 \times 10^{5}$
            cm and $r_{i} = 2 \times 10^{6}$ cm. Deuterium production remains
            considerably large in the trough of $ ^{4}$He production that 
            arises in Figure~\ref{fig:FnlHe4}.}
\end{figure}

\begin{figure}
   \includegraphics[scale=0.80,angle=90]{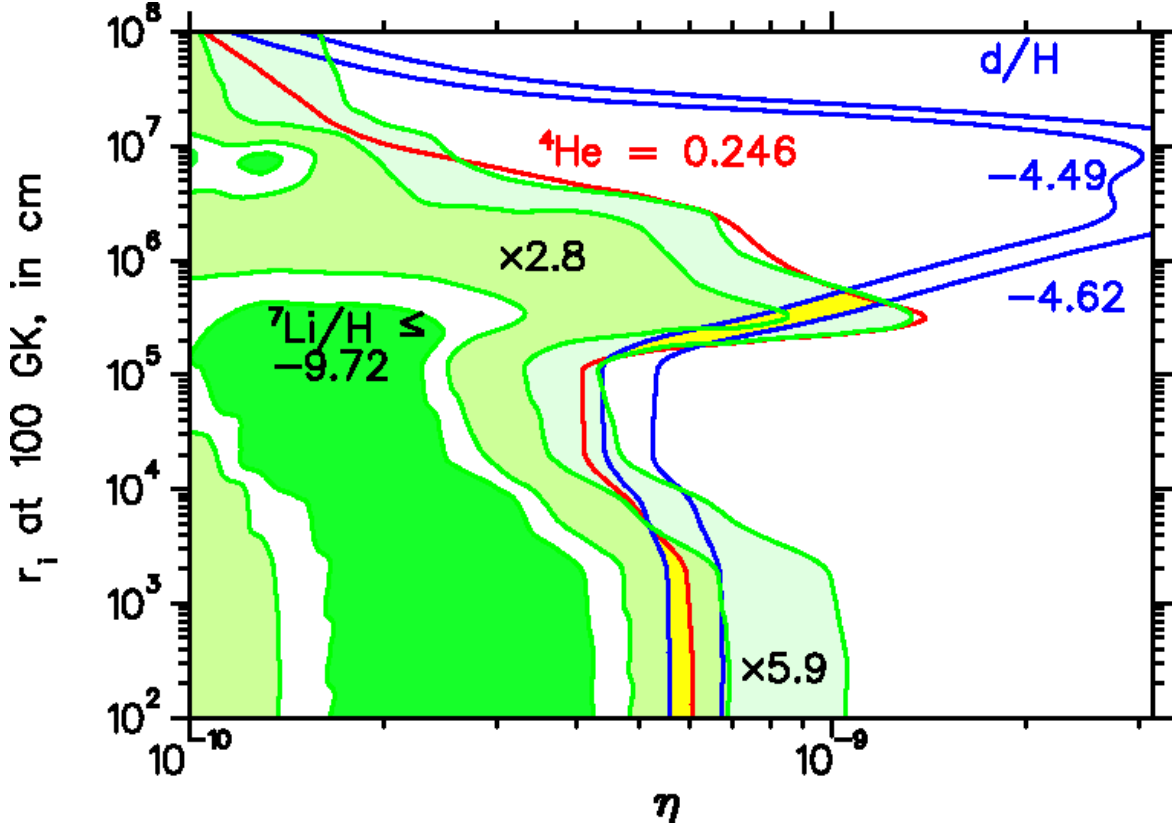}   
   \caption{\label{fig:ibbnocry} ( Color Online ) Observational constraints of
            $ ^{4}$He, deuterium, and $ ^{7}$Li are shown on the IBBN
            cylindrical shell model with $1 - \sqrt{1 - f_{v}}  = $ 0.075, and
            $R_{\rho} = 10^{6}$.  The constraints 
            $Y( ^{7}\li)/Y(\pro) = 1.23_{-0.32}^{+0.68} \times 10^{-10}$ 
            \cite{rbofn00} are shown.  Also shown are $ ^{7}$Li constraints 
            with depletion factors of 2.8 and 5.9.}
\end{figure}

\begin{figure}
   \includegraphics[scale=0.80,angle=90]{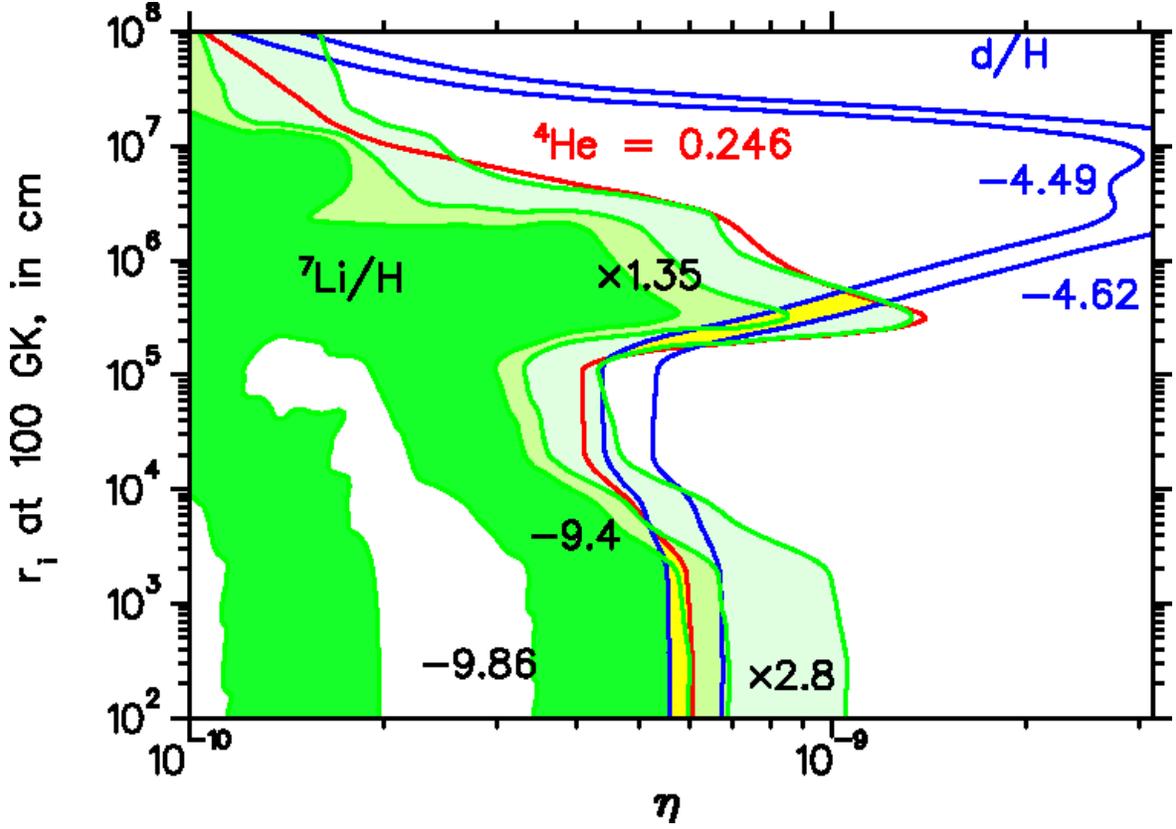}   
   \caption{\label{fig:ibbnocmr} ( Color Online ) Same as in 
            Figure~\ref{fig:ibbnocry}, but showing the constraints 
            $Y( ^{7}\li)/Y(\pro) = 2.34_{-0.96}^{+1.64} \times 10^{-10}$
            \cite{mr04}.  Also shown are these constraints with depletion 
            factors 1.35 and 2.8.}
\end{figure}

For models with the time of diffusion close to the time of nucleosynthesis 
neutron back diffusion becomes less effective.  Nucleosynthesis is concentrated
in the volume fimmediately around the boundary.  This concentration leads to 
decreasing $ ^{4}$He and $ ^{7}$Li+$ ^{7}$Be production, and increasing 
deuterium production.  In models where the time of neutron diffusion coincides
with nucleosynthesis neutrons are not homogenized during nucleosynthesis.  An 
increasing neutron number density remains in the outermost shells as well as a 
decreasing number density in the innermost shells.  $ ^{4}$He, $ ^{7}$Li and 
$ ^{7}$Be production jumps in the high density outermost shells, and overall 
production of these isotopes increases.  But between the boundary and the 
outermost shells are shells with a trough of low $ ^{4}$He production.  
Deuterium is produced in large amounts in that trough.  The deuterium contour 
lines in Figure~\ref{fig:cmdeu} diverge from the contour lines in 
Figures~\ref{fig:cmhe4} and~\ref{fig:cmdeu}.

For models where neutron diffusion peaks at the same time as nucleosynthesis 
the trough in $ ^{4}$He production has disappeared.  Deuterium production 
decreases and the deuterium contour lines in Figure~\ref{fig:cmdeu} are in 
line with the lines in Figures~\ref{fig:cmhe4} and~\ref{fig:cmdeu}.  The 
divergence in the directions of contour lines is significant in setting 
constraints on $\eta$ and $r_{i}$ in the IBBN model.

Application of observational constraints to this IBBN model found slivers of 
concordance between the most recent deuterium constraints \cite{ktsol03} and
$ ^{4}$He constraints by IT04 \cite{it04}.  Concordance occurs for 
$\eta = ( 5.6 - 6.1 ) \times 10^{-10}$ and $r_{i} \le $ 5000 cm, and for 
$\eta = ( 4.3 - 12.3 ) \times 10^{-10}$ and 
$r_{i} = ( 1.3 - 6.0 ) \times 10^{5}$ cm.  The point of divergence between the 
$ ^{4}$He and deuterium contour lines sets the maximum limit of acceptable
$\eta$.  The reliability of $ ^{4}$He constraints remains controversial
\cite{os04}.

Contour lines between $ ^{4}$He, deuterium, and $ ^{7}$Li run roughly 
parallel to each other.  The region Figure~\ref{fig:ibbnocry} marked by the 
$ ^{7}$Li constraints by Ryan et al \cite{rbofn00} then does not have an 
overlap with the slivers of concordance of $ ^{4}$He and deuterium.  A 
depletion factor of 2.8 would bring concordance in both the cases of SBBN and 
the first region of concordance.  But because of the larger shift of the 
$ ^{7}$Li contour lines to lower $\eta$ a larger depletion factor of 5.9 is 
needed to bring the $ ^{7}$Li constraints in agreement with the second region 
of concordance.  Recent $ ^{7}$Li constraints by Melendez \& Ramirez 
\cite{mr04} have weak concordance with $ ^{4}$He and deuterium constraints in
the SBBN case.  But an IBBN model still allows for a larger range of depletion
factor, up to 2.8, to have $ ^{7}$Li be in concordance with $ ^{4}$He and
deuterium.

The IBBN abundance results for $ ^{7}$Li will be compared with new measurements
of the $ ^{7}$Li primordial abundance derived from the ratio 
$(  ^{7}\li/^{6}\li )$ measured in the InterStellar Medium 
\cite{ksak03,kawanomoto05}.  A new neutron lifetime 
$\tau_{\neu} = 878.5 \pm 0.7 \pm$ 0.3 seconds has recently been measured 
\cite{serebrov05}.  Constraints on $\eta$ in an SBBN model have been reassessed
with the new lifetime \cite{mks05}, and the constraints on $\eta$ and $r_{i}$ 
in Figures~\ref{fig:ibbnocry} and~\ref{fig:ibbnocmr} will also be reassessed 
with the new lifetime in an upcoming article \cite{lara05}.  Additionally, this
article will be followed up by articles applying an original solution of the 
neutrino heating effect \cite{lara01a} to both SBBN and IBBN models.  

\section{Acknowledgements}

This work was partially funded by National Science Foundation grants PHY 
9800725, PHY 0102204, and PHY 035482.  This author thanks the Center for 
Relativity at the University of Texas at Austin for the opportunity to work on 
this research, and  Professor Richard Matzner in particular for his help in 
preparing this article.  This author also thanks Professor Toshitaka Kajino of 
the National Astronomical Observatory of Japan for his advice on what to focus 
in this article.

\appendix

\section{Trace of the Texas IBBN Code}

\label{app:trace}
The stretching function that sets the radii of the zones in the cylindrical 
shell model

\begin{eqnarray}
   \xi(r) & = & \xi(r_{b}) + \frac{1}{C_{1}} 
                \left ( 1 - \frac{1}{C_{3}} \right )
                \sqrt{\frac{C_{2}}{C_{3}}} \arctan
                \left [ ( r - r_{b} ) \sqrt{\frac{C_{3}}{C_{2}}} \right ] +
                \frac{r - r_{b}}{C_{1} C_{2}}
   \label{eq:xi}
\end{eqnarray}

\noindent was used by KM90 \cite{km90}.  The radius $r$ is normalized to equal 
64 for the full radius of the model.  Eq.~\ref{eq:xi} maps radii $r$ of the 
zone boundaries to unit values of $\xi$.  Zones near $r_{b}$ have a width 
around $C_{1}$.  Zones far from $r_{b}$ have widths determined by $C_{3}$ and a
rate of zone-size change controlled by $C_{2}$.  $r_{b}$ always corresponds to 
a unit value of $\xi$.  For the model in this article there are 20 zones 
covering the high density outer shell and 44 covering the low density inner 
region.  The baryon number densities $n_{b-high}$ in the outer shell's zones 
and $n_{b-low}$ in the inner region's zones are set

\begin{eqnarray}
    n_{b-low} & = & \frac{n_{b0}}{f_{v}R_{\rho} + ( 1 - f_{v} )} \\
   n_{b-high} & = & R_{\rho} n_{b-low}
\end{eqnarray}

\noindent such that the number density averages out to $n_{b0}$ over the whole
model.

At any given timestep the code solves the differential equation \cite{mmaf90}

\begin{eqnarray}
   \frac{\partial n(i,s)}{\partial t} & = & n_{b}(s) \sum_{j,k,l} N_{i} \left
         ( -\frac{Y^{N_{i}}(i,s) Y^{N_{j}}(j,s)}{N_{i}!N_{j}!}[ij] + 
          \frac{Y^{N_{k}}(k,s) Y^{N_{l}}(l,s)}{N_{k}!N_{l}!}[kl] \right ) 
         \nonumber  \\
   &   & -3\dot{\alpha_{R}}n(i,s) + \frac{1}{r^{p}}\frac{\partial}{\partial r}
           \left ( r^{p} D_{n} \frac{\partial \xi}{\partial r}
                   \frac{\partial n(i,s)}{\partial \xi} \right )
   \label{eq:dndt}
\end{eqnarray}

\noindent for the number density $n(i,s)$ of isotope $i$ in zone $s$.  The 
first two terms correspond to the weak and nuclear reactions that destroy 
( $[ij]$ ) or create ( $[kl]$ ) isotope $i$ within zone $s$.  $n_{b}(s)$ is 
the total baryon number density in zone $s$ and $Y(i,s)$ is the abundance 
$Y(i,s)$ of isotope $i$ in zone $s$.

\begin{eqnarray*}
   Y(i,s) & = & \frac{n(i,s)}{n_{b}(s)}
\end{eqnarray*}

\noindent The $3 \dot{\alpha_{R}} n(i,s)$ term corresponds for the expansion of
the universe, where $R$ is the expansion coefficient of the universe and 
$\alpha_{R} = \ln R$.  This term can be eliminated by transforming to comoving
coordinates.  From here on $r$ will be in comoving coordinates.  The last term 
corresponds to diffusion of isotope $i$ into and out of zone $s$.  The factor 
$p$ depends on the geometry of the model.  $p = $ 0 for planar symmetry, 1 for
cylindrical symmetry and 2 for spherical symmetry.  Currently only neutrons can
diffuse in the Texas IBBN code.  The neutron diffusion coefficient $D_{\neu}$
is calculated from the coefficients $D_{\neu\ele}$ for neutron electron 
scattering and $D_{\neu\pro}$ for neutron proton scattering.

\begin{eqnarray}
   \frac{1}{D_{\neu}} & = & \frac{1}{D_{\neu\ele}} + \frac{1}{D_{\neu\pro}}
\end{eqnarray}

\noindent Banerjee and Chitre \cite{bc91} derived a master equation for the 
diffusion coefficient between two particles scattering off of each other, based
on the first order Chapman-Enskog approximation \cite{glw80}.  Kurki-Suonio et
al (KAGMBCS92) \cite{kagmbcs92}, and Jedamzik and Rehm \cite{jr01} derive the 
same equation for the diffusion coefficient $D_{\neu\ele}$ for neutron-electron
scattering.

\begin{eqnarray}
   D_{\neu\ele} & = & \frac{3}{8} \sqrt{\frac{\pi}{2}} \frac{c}{n_{\ele}
                      \sigma_{\neu\ele}} \frac{K_{2}(z)}{\sqrt{z} K_{5/2}(z)}
                      ( 1 - \frac{n_{\neu}}{n_{t}} )
\end{eqnarray}

\noindent $K_{2}(z)$ and $K_{5/2}(z)$ are modified Bessel functions of order 2 
and 5/2, $\sigma_{\neu\ele}$ is the transport cross section of the scattering 
and $z = m_{\ele}/ kT$.  $n_{\neu}/n_{t}$ is the neutron fraction of the total 
number of ALL particles.  This fraction is of the order $10^{-10}$ and so can 
be ignored.  For neutron-proton scattering Jedamzik and Rehm \cite{jr01} derive
an updated expression for the diffusion coefficient $D_{\neu\pro}$ 

\begin{eqnarray}
   D_{\neu \pro} & = & \frac{3}{8 \sqrt{\pi}} \frac{c}{a_{s}^{2}} 
                       \frac{1}{n_{\pro}} \sqrt{\frac{k_{B} T}{m_{N} c^{2}}} 
     \frac{1}{I(a_{1},b_{1}) + \frac{3 a_{t}^{2}}{a_{s}^{2}} I(a_{2},b_{2})} \\
          I(a,b) & = & \frac{1}{2} \int_{0}^{\infty} dx \frac{x^{2} e^{-x}}
                      {ax + \left ( 1 - \frac{bx}{2} \right )^{2}} \nonumber \\
           a_{1} & = & a_{s}^2 \frac{m_{N} c^{2}}{\hbar^{2} c^{2}} k_{b} T 
                       \nonumber \\
           b_{1} & = & r_{s} a_{s} \frac{m_{N} c^{2}}{\hbar^{2} c^{2}} k_{b} T 
                       \nonumber \\
           a_{2} & = & a_{t}^2 \frac{m_{N} c^{2}}{\hbar^{2} c^{2}} k_{b} T 
                       \nonumber \\
           b_{2} & = & r_{t} a_{t} \frac{m_{N} c^{2}}{\hbar^{2} c^{2}} k_{b} T
                       \nonumber
\end{eqnarray}

\noindent $m_{N}$ is the nucleon mass.  The parameters $a_{s} = $ -23.71 fm, 
$r_{s} = $ 2.73 fm, $a_{t} = $ 5.432 fm, and $r_{t} = $ 1.749 fm come from 
singlet and triplet scattering.

The Texas IBBN code progresses a timestep $\Delta t_{m}$ for each step $m$.  
Eq.~\ref{eq:dndt} is evolved using a implicit second order Runge-Kutta method
\cite{kawano92}.  To use this method, Eq.~\ref{eq:dndt} has to be linearized.
The weak-nuclear reaction terms can be linearized in a manner similar to the 
linearization of abundances $Y$ used by Wagoner \cite{wagoner69} and this 
author \cite{lara98}.

\begin{eqnarray}
   \frac{n_{m A}(i,s) - n_{m-1}(i,s)}{\Delta t_{m-1}} 
 & = & \sum_{j,k,l} - \frac{N_{i} [ij]}{N_{i}! N_{j}! ( N_{i} + N_{j} )}
     [ N_{i}Y_{m}^{N_{i}-1}(i,s) Y_{m}^{N_{j}}(j,s) n_{m A}(i,s) + \nonumber \\
 &   & N_{j}Y_{m}^{N_{i}}(i,s) Y_{m}^{N_{j}-1}(j,s) n_{m A}(j,s) ] \nonumber \\
 &   & + \frac{N_{i} [kl]}{N_{k}! N_{l}! ( N_{k} + N_{l} )} 
     [ N_{k}Y_{m}^{N_{k}-1}(k,s) Y_{m}^{N_{l}}(l,s) n_{m A}(k,s) + \nonumber \\
 &   & N_{l}Y_{m}^{N_{k}}(k,s) Y_{m}^{N_{l}-1}(l,s) n_{m A}(l,s) ] + \cdots
   \label{eq:rkA1}
\end{eqnarray}

\noindent $\Delta t_{m-1}$ is the time difference between step $m - 1$ and step
$m$.  For the diffusion term the zones are defined on a grid whose points 
$r(s)$ correspond to the outer radii of zones $s$.  Number densities $n(i,s)$ 
are considered the number densities at the midpoint radius between the inner
and outer radii of zone $s$.  The points $r(s)$ correspond to points in 
$\xi (s)$ space a distance of one unit between each other.  The first space 
derivative in the diffusion term in Eq.~\ref{eq:dndt} can be discretized as:

\begin{eqnarray*}
   \frac{\partial n}{\partial t} & = & \cdots \frac{1}{r^{p}} 
    \frac{\partial}{\partial r} \left ( r^{p} D \frac{\partial \xi}{\partial r}
                                    \frac{\partial n}{\partial \xi} \right ) \\
   \frac{\partial n}{\partial t} & = & \cdots \frac{1}{r^{p}} 
                                       \frac{\partial}{\partial r} \left [ 
               \left ( r^{p} D \frac{\partial \xi}{\partial r} \right )_{s-1/2}
                                       \frac{n[r(s)] - n[r(s-1)]}{1} \right ]
\end{eqnarray*}

\noindent Note that the coefficient $[ r^{p} D (\partial \xi)/(\partial r) ]$ 
depends on $r$.  The $(1/r^{p}) (\partial / \partial r)$ can be rewritten as a
partial derivative of $r^{p+1}$.  One can then write the discretization of the
second space derivative as:

\begin{eqnarray*}
   \frac{\partial n}{\partial t} & = & \cdots ( p + 1 )
                                   \frac{\partial}{\partial (r^{p+1})} \left [ 
               \left ( r^{p} D \frac{\partial \xi}{\partial r} \right )_{s-1/2}
                                      \frac{n[r(s)] - n[r(s-1)]}{1} \right ] \\
   \frac{\partial n}{\partial t} & = & \cdots ( p + 1 ) \left (
             \frac{\left ( r^{p} D \frac{\partial \xi}{\partial r} \right )_{s}
                 \{ n[r(s + \frac{1}{2})] - n[r(s - \frac{1}{2})] \}} 
                                       {r^{p+1}(s) - r^{p+1}(s-1)} \right. \\
                                 &   & \left. - \frac{
                 \left ( r^{p} D \frac{\partial \xi}{\partial r} \right )_{s-1}
                           \{ n[r(s - \frac{1}{2})] - n[r(s - \frac{3}{2})] \}}
                                       {r^{p+1}(s) - r^{p+1}(s-1)} \right )
\end{eqnarray*}

\noindent For an isotope $i$ the densities $n[r(s + 1/2)]$, $n[r(s - 1/2)]$ and
$n[r(s - 3/2)]$ are defined as $n(i,s+1)$, $n(i,s)$ and $n(i,s-1)$ 
respectively.  The coefficient $[ r^{2} D (\partial \xi)/(\partial r) ]_{s}$ is
calculated at $r(s)$.  One can apply this discretization to an implicit version
of the diffision part of Eq.~\ref{eq:dndt}.  

\begin{eqnarray}
   \frac{n_{m A}(i,s) - n_{m-1}(i,s)}{\Delta t_{m-1}} 
    & = & \cdots - \frac{( p + 1 ) \left 
          ( r^{p} D \frac{\partial \xi}{\partial r} \right )_{s}}
          {r^{p+1}(s) - r^{p+1}(s-1)} n_{m A}(i,s+1) \nonumber \\
    &   & + ( p + 1 ) \frac{\left ( r^{p} D \frac{\partial \xi}{\partial r} 
          \right )_{s} + \left ( r^{p} D \frac{\partial \xi}{\partial r} 
          \right )_{s-1}}{r^{p+1}(s) - r^{p+1}(s-1)} n_{m A}(i,s) \nonumber \\
    &   & - \frac{( p + 1 ) \left ( r^{p} D \frac{\partial \xi}{\partial r} 
          \right )_{s-1}}{r^{p+1}(s) - r^{p+1}(s-1)} n_{m A}(i,s-1)
   \label{eq:rkA2}
\end{eqnarray}

\noindent Any baryons that flow out beyond the distance scale are assumed to be
replenished by baryons flowing in from other sets of shells, and $r(0) = $ 0 is
the center of the shells.  The code uses reflective boundary conditions at the
endpoints of the grid.

\begin{eqnarray*}
   \frac{n_{m A}(i,1)  - n_{m-1}(i,1)}{\Delta t_{m-1}} 
    & = & \cdots - \frac{( p + 1 ) \left 
          ( r^{p} D \frac{\partial \xi}{\partial r} \right )_{1}}
          {r^{p+1}(1) - r^{p+1}(0)} n_{m A}(i,2) \\
    &   & + \frac{( p + 1 ) \left 
          ( r^{p} D \frac{\partial \xi}{\partial r} \right )_{1}}
            {r^{p+1}(1) - r^{p+1}(0)} n_{m A}(i,1) \\
   \frac{n_{m A}(i,64)  - n_{m-1}(i,64)}{\Delta t_{m-1}} 
    & = & \cdots + \frac{( p + 1 ) \left 
          ( r^{p} D \frac{\partial \xi}{\partial r} \right )_{64}}
          {r^{p+1}(64) - r^{p+1}(63)} n_{m A}(i,64) \\
    &   & - \frac{( p + 1 ) \left 
          ( r^{p} D \frac{\partial \xi}{\partial r} \right )_{63}}
          {r^{p+1}(64) - r^{p+1}(63)} n_{m A}(i,63)
\end{eqnarray*}

\noindent where $r(64) = $ distance scale $r_{i}$.  The above equations can be
applied to the diffusion of any isotope $i$, but only neutrons ( $i = $ 1 )
diffuse for the results of this article. 

Eq.~\ref{eq:rkA1} and Eq.~\ref{eq:rkA2} combined together can be rewritten as a
matrix equation for a new number density value $n_{m A}(i,s)$.  The matrix 
consists of a 68 $\times$ 68 matrix for each of the 64 zones, built from the 
terms in Eq.~\ref{eq:rkA1}.  From Eq.~\ref{eq:rkA2} come terms that couple 
$n(1,s)$ with $n(1,s+1)$ and $n(1,s-1)$ due to neutron diffusion.  
$n_{m A}(i,s)$ is then used in the following equation

\begin{eqnarray*}
 \tilde{n}_{m}(i,s) & = & n_{m}(i,s) + \left [ 
                          \frac{n_{m A}(i,s) - n_{m-1}(i,s)}{\Delta t_{m-1}}
                          \right ] \Delta t_{m}
\end{eqnarray*}

\noindent to calculate an interim value $\tilde{n}_{m}(i,s)$ of the number 
densities.  This is the first step of the Runge-Kutta method, with 
$\tilde{n}_{m}(i,s)$ the first estimate of the values of $n_{m}(i,s)$ at time
$t_{m} + \Delta t_{m}$.  Using    
$\tilde{Y}_m(i,s) = \tilde{n}_{m}(i,s) / \tilde{n}_{b}(s)$ the code solves a 
second matrix equation

\begin{eqnarray*}
   \frac{n_{m B}(i,s) - n_{m}(i,s)}{\Delta t_{m}} 
        & = & \sum_{j,k,l} - \frac{N_{i} [ij]}{N_{i}! N_{j}! ( N_{i} + N_{j} )}
              [ N_{i}\tilde{Y}_{m}^{N_{i}-1}(i,s) \tilde{Y}_{m}^{N_{j}}(j,s) 
              n_{m B}(i,s) + \\
        &   & N_{j}\tilde{Y}_{m}^{N_{i}}(i,s) \tilde{Y}_{m}^{N_{j}-1}(j,s) 
              n_{m B}(j,s) ] \\
        &   & + \frac{N_{i} [kl]}{N_{k}! N_{l}! ( N_{k} + N_{l} )} 
              [ N_{k}\tilde{Y}_{m}^{N_{k}-1}(k,s) \tilde{Y}_{m}^{N_{l}}(l,s) 
              n_{m B}(k,s) + \\
        &   & N_{l}\tilde{Y}_{m}^{N_{k}}(k,s) \tilde{Y}_{m}^{N_{l}-1}(l,s) 
              n_{m B}(l,s) ] - \\
        &   & \frac{( p + 1 ) \left 
              ( r^{p} D \frac{\partial \xi}{\partial r} \right )_{s}}
              {r^{p+1}(s) - r^{p+1}(s-1)} n_{m B}(i,s+1) \\
        &   & + ( p + 1 ) \frac{\left ( r^{p} D \frac{\partial \xi}{\partial r}
              \right )_{s} + \left ( r^{p} D \frac{\partial \xi}{\partial r} 
              \right )_{s-1}}{r^{p+1}(s) - r^{p+1}(s-1)} n_{m B}(i,s) \\
        &   & - \frac{( p + 1 ) \left ( r^{p} D \frac{\partial \xi}{\partial r}
              \right )_{s-1}}{r^{p+1}(s) - r^{p+1}(s-1)} n_{m B}(i,s-1)
\end{eqnarray*}

\noindent for new number density values $n_{m B}(i,s)$.  $\Delta t_{m}$ is the
time difference between step $m$ and step $m + 1$.  Final new values for 
$n_{m+1}(i,s)$ at timestep $m + 1$ can then be calculated.

\begin{eqnarray}
   n_{m+1}(i,s) & = & n_{m}(i,s) + \frac{1}{2} \left [ 
                     \frac{n_{m A}(i,s) - n_{m-1}(i,s)}{\Delta t_{m-1}} +
                     \frac{n_{m B}(i,s) - n_{m}(i,s)}{\Delta t_{m}} \right ]
                     \Delta t_{m}
\end{eqnarray}

\noindent This is the second step ( ``$B$'' ) of the Runge-Kutta method.  At 
the same time as with $n_{m}(i,s)$ the Texas IBBN evolves $\ln R$ and the 
electromagnetic plasma energy density $\rho_{\ele+\gamma}$  

\begin{eqnarray}
             \frac{d ( \ln R )}{dt} & = & \sqrt{\frac{8}{3} \pi G 
                               ( \rho_{\gamma} + \rho_{\ele} + \rho_{\nu} )} \\
   \frac{d\rho_{\ele + \gamma}}{dt} & = & - 4 \frac{\dot{R}}{R} \rho_{\gamma} -
                                 3 \frac{\dot{R}}{R} ( p_{\ele} + \rho_{\ele} )
\end{eqnarray}

\noindent also by the Runge-Kutta method.

After both Runge-Kutta steps have been done the code determines the new baryon 
number density $n_{b}(s)$ of each zone using 

\begin{eqnarray*}
   n_{b}(s) & = & \sum_{i = 1}^{68} A_{i} n(i,s)
\end{eqnarray*}

\noindent where $A_{i}$ is the atomic weight of isotope $i$.  From $n_{b}(s)$ 
and $n(i,s)$ the code can calculate $Y(i,s)$.   At any given time the abundance
$Y_{av}(i)$ and mass fraction $X_{i}$ of isotope $i$ in the entire model can be
calculated from $Y(i,s)$ using

\begin{eqnarray*}
   Y_{av}(i) & = & \frac{\sum_{s = 1}^{64} n(i,s)[ r^{p+1}(s) - r^{p+1}(s-1) ]}
                  {\sum_{s = 1}^{64} n_{b}(s) [ r^{p+1}(s) - r^{p+1}(s-1) ]} \\
       X_{i} & = & A_{i} Y_{av}(i)
\end{eqnarray*}

\noindent These overall abundances and mass fractions can be shown as contour
maps of the IBBN code's parameters, and compared to observational constraints.

\bibliography{PRD01}

\clearpage

\clearpage

\clearpage

\clearpage

\clearpage

\clearpage

\clearpage

\clearpage

\clearpage

\clearpage

\clearpage

\clearpage

\end{document}